\documentclass  [twocolumn, amsmath,amssymb,aps,prl,showpacs, superscriptaddress,longbibliography, nobibnotes]{revtex4-1}

\usepackage{graphicx}
\usepackage{dcolumn}
\usepackage{bm}
\usepackage{braket,amsmath,amssymb,bm}
\usepackage{color}
\usepackage{siunitx}
\usepackage{float}
\usepackage{hyperref}
\usepackage{lineno}
\usepackage{multirow}
\usepackage{tabularx}
\usepackage{enumitem}
\usepackage[version=3]{mhchem} 
\usepackage{braket}
\usepackage{siunitx} 
\usepackage{color}
\usepackage{bm}
\usepackage{lineno}
\usepackage{braket,amsmath,amssymb,bm}
\floatstyle{plaintop}
\restylefloat{scheme}
\floatstyle{plain}

\begin{document}

\title{\texorpdfstring{Quantum-grade nanodiamonds for ultrabright spin detection in live cells}{Quantum-grade nanodiamonds for ultrabright spin detection in live cells}}

\author{Keisuke Oshimi}
\affiliation{Department of Chemistry, Graduate School of Life, Environmental, Natural Science and Technology, Okayama University, Okayama 700-8530, Japan}
\author{Hiromu Nakashima}
\affiliation{Department of Chemistry, Graduate School of Life, Environmental, Natural Science and Technology, Okayama University, Okayama 700-8530, Japan}
\author{Sara Mandi\'{c}}
\affiliation{Department of Chemistry, Graduate School of Life, Environmental, Natural Science and Technology, Okayama University, Okayama 700-8530, Japan}
\author{Hina Kobayashi}
\affiliation{Department of Chemistry, Graduate School of Life, Environmental, Natural Science and Technology, Okayama University, Okayama 700-8530, Japan}
\author{Minori Teramoto}
\affiliation{Advanced Materials Laboratory, Sumitomo Electric Industries, Ltd., Hyogo 664-0016, Japan}
\author{Hirokazu Tsuji}
\affiliation{Advanced Materials Laboratory, Sumitomo Electric Industries, Ltd., Hyogo 664-0016, Japan}
\author{Yoshiki Nishibayashi}
\affiliation{Advanced Materials Laboratory, Sumitomo Electric Industries, Ltd., Hyogo 664-0016, Japan}
\author{Yutaka Shikano}
\affiliation{Institute of Systems and Information Engineering, University of Tsukuba, Tsukuba, Ibaraki 305-8573, Japan}
\affiliation{Center for Artificial Intelligence Research (C-AIR), University of Tsukuba, Tsukuba, Ibaraki 305-8577, Japan}
\affiliation{Institute for Quantum Studies, Chapman University, Orange, CA 92866, USA}
\author{Toshu An}
\affiliation{School of Materials Science, Japan Advanced Institute of Science and Technology, Nomi, Ishikawa 923-1292, Japan}
\author{Masazumi Fujiwara}
\email{masazumi@okayama-u.ac.jp}
\affiliation{Department of Chemistry, Graduate School of Life, Environmental, Natural Science and Technology, Okayama University, Okayama 700-8530, Japan}

\begin{abstract}
Optically accessible spin-active nanomaterials are promising as quantum nanosensors for probing biological samples. 
However, achieving bioimaging-level brightness and high-quality spin properties for these materials is challenging and hinders their application in quantum biosensing.
Here, we demonstrate ultrabright fluorescent nanodiamonds (NDs) containing 0.6--1.3-ppm nitrogen-vacancy (NV) centers by spin-environment engineering via enriching spin-less \ce{^12C}-carbon isotopes and reducing substitutional nitrogen spin impurities. 
The NDs, readily introduced into cultured cells, exhibited substantially narrow optically detected magnetic resonance (ODMR) spectra, requiring 16-times less microwave excitation power to give an ODMR depth comparable to that of conventional type-Ib NDs.
They show average spin-relaxation times of $T_1 = 0.68\,\si{\ms}$ and $T_2 = 1.6\,\si{\us}$ (1.6 ms and 2.7 \si{\us} maximum) that were 5- and 11-fold longer than those of type-Ib, respectively.
The bulk-like NV spin properties and bright fluorescence demonstrated in this study significantly improve the sensitivity of ND-based quantum sensors for biological applications.

\end{abstract}

\maketitle

\section{Introduction}
Engineering spin-active materials is critical to developing highly sensitive quantum nanosensors, as demonstrated by nanodiamonds (NDs) containing color defect centers~\cite{kucsko2013nanometre, miller2020spin}, organic nanosolids with radical molecules~\cite{geng2023sub}, rare earth nanocrystals~\cite{liu2020defect}, and nanoflakes of hexagonal boron nitrides~\cite{ronceray2023liquid, aharonovich2022quantum}. 
Among these spin-active materials, NDs containing nitrogen-vacancy (NV) centers are the most advanced quantum nanosensors utilized for biological applications owing to their multimodal sensing capability~\cite{aslam2023quantum, knowles2multimodal}, photostability~\cite{haziza2017fluorescent}, chemical functionality~\cite{komatsu2023poly}, and biocompatibility~\cite{vaijayanthimala2012long}. 
The NV quantum nanosensors exploit the dependence of optically accessible NV electron spins on magnetic field, electric field, and temperature of the surrounding cellular environment~\cite{doherty2013nitrogen}, which enables sub-cellular measurements of either those values or extended physicochemical parameters, such as pH~\cite{fujisaku2019ph, sow2020high}, magnetic ions~\cite{steinert2013magnetic, freire2023role}, reactive oxygen species~\cite{nie2021quantum,sigaeva2022diamond} and rheology~\cite{feng2021association, gu2023simultaneous}. 
In the majority of the cases, these sensing modalities measure the frequency shift (or modulated relaxation times) in optically detected magnetic resonance (ODMR) of NV centers. 
Therefore, measurement sensitivity is critically dependent on the quality of the NV spin properties~\cite{barry2020sensitivity}.

However, current fluorescent NDs incorporating high-density NVs that show bioimaging-level brightness exhibit deficient spin qualities in contrast to bulk diamonds. 
They show broad ODMR spectra and short spin relaxation times, which substantially deteriorates the measurement sensitivity~\cite{barry2020sensitivity, degen2017quantum}. 
The NV spins are affected by (i) a high concentration of spin impurities and (ii) surface spin noise~\cite{mizuochi2009coherence,knowles2014observing, romach2015spectroscopy}. 
However, recent studies suggest that the effects of these factors on NDs might be mitigated.
First, high-quality synthetic bulk diamonds with dense NV ensembles have realized high-yield NV production while enriching spinless \ce{^12C}-carbon isotopes and reducing substitutional nitrogen impurity for quantum sensor applications ~\cite{andrich2014engineered,herbschleb2019ultra, bauch2020decoherence, shinei2021equilibrium}. 
These bulk diamonds exhibited high NV spin qualities and demonstrated fascinating applications, including nanoscale nuclear magnetic resonance~\cite{wolf2015subpicotesla, abobeih2019atomic} and neural monitoring~\cite{aslam2023quantum,zhang2023optimizing}.
Second, previous studies revealed that the surface spin noises reduced the NV spin relaxation time in NDs with $d^{-4}$ scaling for the ND diameter ($d$)~\cite{tetienne2013spin}. 
The spin deterioration of NVs is prominent only for $d < 80$ nm, and this surface effect might not be predominant for a majority of the NDs used in the quantum biological applications owing to their relatively large diameter ($d > 80$ nm)~\cite{kucsko2013nanometre,miller2020spin,fujiwara2020real, feng2021association}, which is corroborated by observation of single-NV NDs with long spin coherence, wherein the spin impurities could be minimized to ppb level for quantum applications requiring single NVs~\cite{knowles2014observing, march2023long}.

Here, we demonstrate NDs possessing bulk-like NV spin properties by controlling the spin impurities of \ce{^13C} and N, while increasing the NV concentration for bioimaging-level brightness.
The spin relaxation times of the NVs, i.e., $T_1$ and $T_2$, in these spin-controlled NDs are enhanced by a factor of 5 for $T_1$ and 10 for $T_2$, compared with conventional type-Ib NDs.
The results indicate the possibility of realizing quantum-grade NDs that can implement various quantum-enhanced measurement protocols for biological samples and applications.

\section{Results}
\label{sec3}

\begin{figure*}[th!]
    \centering
    \includegraphics[width=17 cm]{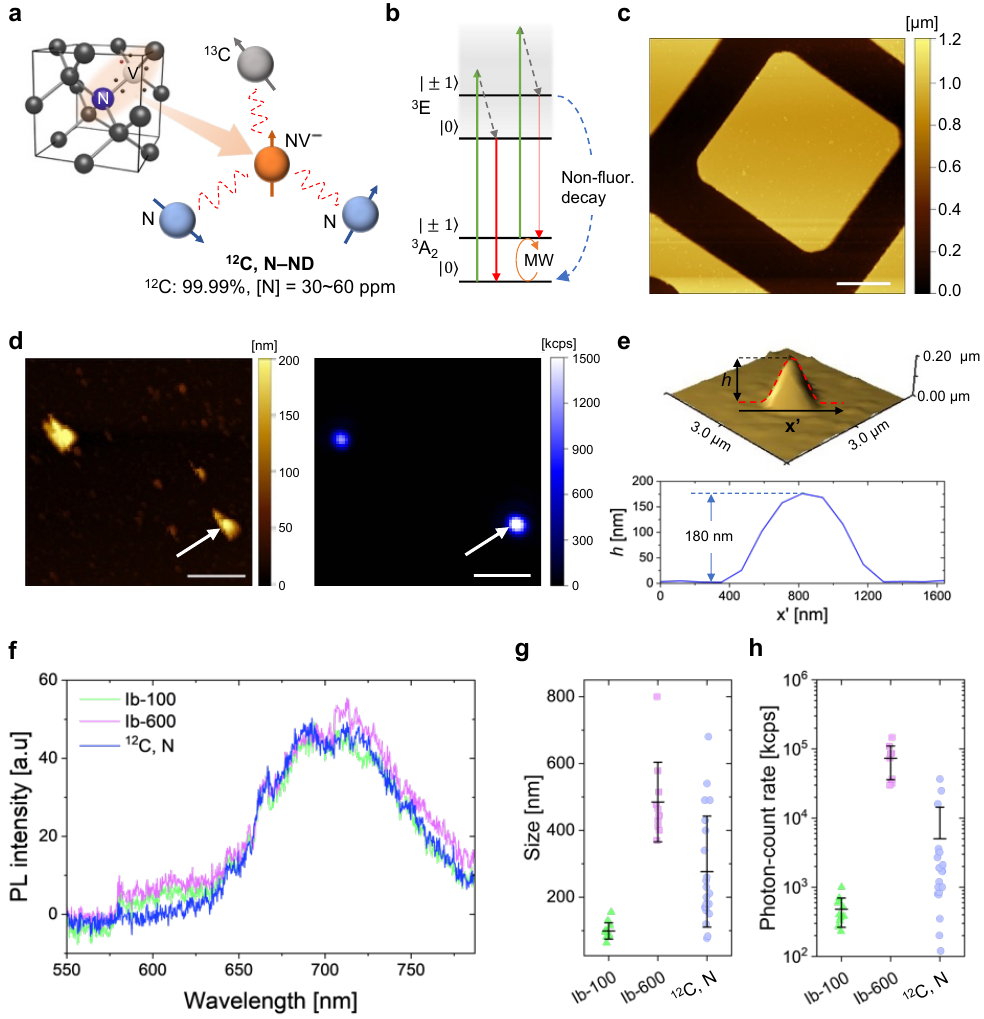}
    \caption{
    (a) Illustrations of the NV crystal structure and the interaction of NV with the spin bath of \ce{N} and \ce{^13C}. 
    (b) Schematic representation of simplified energy level structure of NV centers. $\ket{0}$ and $\ket{\pm 1}$ are the spin sublevels for $m_s=0$ and $m_s=\pm 1$, respectively. MW: microwave. \ce{^3A_2} (\ce{^3E}): triplet ground (excited) state.
    (c) AFM topography image of a single grid engraved on a coverslip. Scale bar: 10\,\si{\um}. 
    (d) AFM topography and the corresponding confocal fluorescence images of \ce{^12C}, N-NDs on a grid. Scale bars: 2\,\si{\um}. 
    (e) Three-dimensional visualization of the topography of the ND indicated by the white arrow in Fig.~\ref{fig1}(d) (top) with a cross-section along the x' axis (bottom). 
    (f) Fluorescent spectra of type-Ib NDs and \ce{^12C}, N-NDs.
    (g) Statistical plots of the ND size determined by AFM and (h) photon-count rate at an optical excitation intensity of $\sim 7\,\si{kW.cm^{-2}}$ for Ib-100, Ib-600, and \ce{^12C}, N-NDs. 
    Mean and standard deviation (1$\sigma$) are indicated in the statistical plots.
    The error bar is shown only for the upper error side ($+\sigma$) for \ce{^12C}, N-NDs in Fig.~\ref{fig1}(h), where a large standard deviation (sigma = 5000 kcps) makes a negative lower side ($-\sigma$) invisible in a log plot.
    }
    \label{fig1}
\end{figure*}

To achieve a high NV concentration exhibiting bioimaging-level brightness with simultaneous reduction in the amount of spin impurities, we use NV-enriched single-crystalline bulk diamonds, wherein the major spin impurities of \ce{N} (the so-called \ce{P_1} center) and \ce{^13C} are minimized (Fig.~\ref{fig1}\textbf{a}). 
We prepared NDs via pulverization of single-crystalline bulk diamonds grown using a high-pressure high-temperature (HPHT) method with controlled impurities of [\ce{^12C}] = 99.99\% and [N] = 30--60 ppm (see Methods).
These NDs (hereafter called \ce{^12C}, N-NDs)  exhibit NV red fluorescence under green optical excitation with ODMR signal upon microwave irradiation (Fig.~\ref{fig1}\textbf{b}).
To characterize the fluorescence brightness and size, the NDs were suspended in water and drop-casted onto coverslips with engraved island grids (Fig.~\ref{fig1}\textbf{c}). 
We performed single-particle characterization using confocal fluorescence and atomic force microscopy (AFM). 
Figure~\ref{fig1}\textbf{d} shows the topography and corresponding confocal fluorescence images of the area containing several \ce{^12C}, N-NDs.
We observe bright fluorescence emission from these NDs (with a photon-count rate of 1500 kcps) at an optical excitation intensity of $\sim 7\,\si{kW.cm^{-2}}$. 
Their mean fluorescence intensity is comparable to the type-Ib NDs, sufficient for bioimaging (Supplementary Figs.~\ref{AFM image for Ib}\textbf{a--d}). 
A close view of the ND topography indicated by the white arrow is shown in Fig~\ref{fig1}\textbf{e}, where the size is determined to be $h = 180$ nm by taking the height of the sample from the xy-imaging plane.
Further, Fig.~\ref{fig1}\textbf{f} shows the typical fluorescence spectra with zero phonon line at 637 nm, confirming that the fluorescence of the \ce{^12C}, N-NDs primarily originates from the negatively charged NVs.

The concentration of NV in NDs is typically determined by electron paramagnetic resonance (EPR) spectroscopy~\cite{zvi2023engineering}. 
However, this method requires hundreds of milligram of samples and cannot be employed for the present NDs due to the small amount.
Alternatively, we determined [NV] of the NDs by measuring their size and fluorescence intensity and comparing them with those of two types of type-Ib NDs having mean sizes of 100 nm (Ib-100) and 600 nm (Ib-600) (Figs.~\ref{fig1}\textbf{g, h}). 
The size analysis revealed the ND mean sizes as 98.5, 485, and 277 nm for the Ib-100, Ib-600, and \ce{^12C}, N-NDs, respectively (Fig.~\ref{fig1}\textbf{g}). 
These mean sizes, which were relatively smaller than the supplier's specification, reflected the typical disk-like ND morphology~\cite{eldemrdash2023fluorescent}, which was more prominent in Ib-600.
Because the number of NVs is proportional to the volume of NDs, we assume that the fluorescence intensity is proportional to the ND volume: $I \propto [\rm NV]\textit{V}$, where $I$ and $V$ are the fluorescence photon-count rate and volume of the NDs, respectively. 
To determine [NV] of the \ce{^12C}, N-NDs from [NV] of Ib-100 (Ib-600), we used the relationship: $[\rm NV]_{\rm 12C, N} = \textit{I}_{\rm 12C, N} \textit{I}_{\rm Ib100}^{-1} \textit{V}_{\rm Ib100} \textit{V}_{\rm 12C, N}^{-1} [\rm NV]_{\rm Ib100}$ ($I_{\rm 12C, N} I_{\rm Ib600}^{-1} V_{\rm Ib600} V_{\rm 12C, N}^{-1} [\rm NV]_{\rm Ib600}$). 
[NV] of the \ce{^12C}, N-NDs was 0.6--1.3 ppm, which was determined based on [NV] $\approx$ 3 and 3.5 ppm for the Ib-100 and for Ib-600 NDs, respectively; these [NV] values of the Ib-100 and for Ib-600 NDs were adopted from the specification sheet and were considered based on the conceivable ND morphology in terms of the aspect ratio of the NDs (see Supplementary Information for the details).

\begin{figure*}[th!]
    \centering
    \includegraphics[width=17 cm]{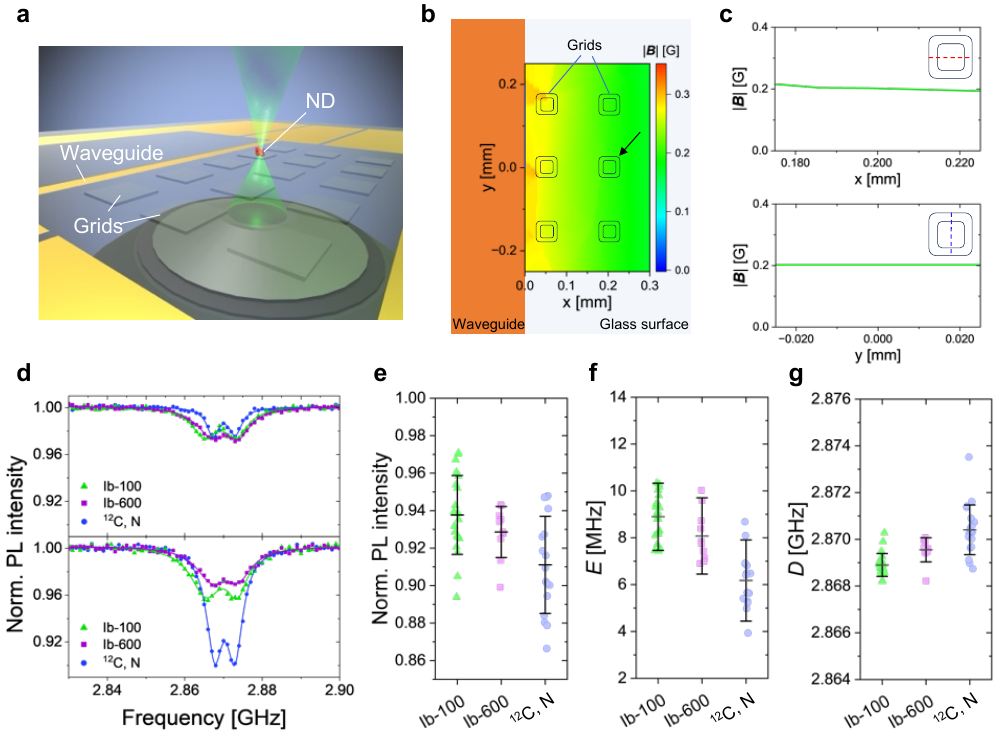}
    \caption{
    (a) Schematic representation of NDs on grids of the notch-shaped microwave antenna. 
    (b) Heat map of simulated magnetic field ($|\boldsymbol{B}|$) on the antenna overlaid with schematic grid structures. 26.1 mW microwave power was used for the simulation.
    (c) Cross-sections of $|\boldsymbol{B}|$ over the grid indicated by a white arrow at $(x, y) = (0.2, 0.0)\,\si{mm}$ along $x$ (top) and $y$ axes (bottom). 
    (d) Representative CW-ODMR spectra of an Ib-100, Ib-600, and \ce{^12C}, N-ND when the microwave power was adjusted to give 3\% ODMR contrast (top) and with the common microwave power of 26.1 mW at the input of the notch area (bottom). 
    A common optical intensity was used in both the panels ($\sim 6\,\si{kW.cm^{-2}}$). 
    The lines are double-Lorentzian fits.
    Statistical plots for the Ib-100 (green triangle), Ib-600 (purple square), and \ce{^12C}, N-NDs (blue circle) for (e) ODMR depth, (f) \textit{E}, and (g) \textit{D}.
    Mean and standard deviation (1$\sigma$) are indicated in the statistical plots.
    }
    \label{fig2}
\end{figure*}

Next, we characterized the NV electron spin properties of the \ce{^12C}, N-NDs by measuring their continuous-wave (CW) ODMR spectra. 
Quantitative characterization of the CW-ODMR spectra over different NDs requires a well-defined spatial microwave-excitation-field pattern because the CW-ODMR spectral shape is sensitive to the applied microwave intensity.
Therefore, we exploited an on-chip platform based on a previously developed notch-shaped antenna pattern~\cite{oshimi2022glass}, which provided a uniform distribution of the microwave magnetic field ($|\bm{B}|$) that was quantitatively defined at each grid (Figs.~\ref{fig2}\textbf{a, b}).

In this experiment, we selected the second nearest grid ($x = 0.2\,\si{mm}$) from the edge of the central waveguide ($x = 0.0\,\si{mm}$), where the variation in $|\bm{B}|$ was only 0.02 G for the microwave power of 26.1 mW (14.2 dBm) utilized in the CW-ODMR experiments, as shown in Figs.~\ref{fig2}\textbf{b, c} (see Table~\ref{tabel1}).
This variation resulted in only a 0.8\% difference in the ODMR depth and a 10\% change in the Rabi-frequency period, which enabled the characterization of the NV spins at the same microwave power without requiring Rabi frequency measurement for each ND.
To observe the ODMR differences among the Ib-100, Ib-600, and \ce{^12C}, N-NDs, the microwave power required to attain a 3\% ODMR contrast was determined for each type of these NDs at the same optical intensity. 
The \ce{^12C}, N-NDs exhibited a 3\% ODMR contrast with a microwave power of $1.04$ mW in the detection area ($|\bm{B}| = 0.041$ G in the grid position), whereas the Ib-100 and Ib-600 NDs required $10.4$ mW ($|\bm{B}| = 0.13$ G) and $20.7$ mW ($|\bm{B}| = 0.18$ G) to obtain a 3\% contrast (top panel in Fig.~\ref{fig2}\textbf{d}). 
This result indicates that to attain the same 3\% ODMR contrast, the microwave power required by the \ce{^12C}, N-NDs is 10 and 20 times lower than those required by the Ib-100 and Ib-600 NDs, respectively. 
Moreover, the ODMR spectra of the \ce{^12C}, N-NDs were substantially narrower than those of the type-Ib NDs.
We determined the spectral evolution of all the NDs under an applied microwave power of 26.1 mW (bottom panel in Fig.~\ref{fig2}\textbf{d}).
The ODMR contrast shown by the \ce{^12C}, N-NDs was considerably deeper than those shown by the Ib-100 and Ib-600 NDs, and this result confirmed the improvement in the NV spin properties of the NDs developed in this study. 

Subsequently, we performed a statistical analysis of the CW-ODMR spectra of these NDs (Figs.~\ref{fig2}\textbf{e--g}).
The results demonstrated that on average, the ODMR depth was greater for the \ce{^12C}, N-NDs than for the Ib-100 and Ib-600 counterparts (Fig.~\ref{fig2}\textbf{e}). 
We further analyzed spectral parameters for peak splitting (\textit{E}) related to crystal strains and zero-field splitting ($D$) to evaluate more NV intrinsic spin properties of the NDs (see Methods).
The mean of \textit{E} for the \ce{^12C}, N-NDs was smaller than that for the type-Ib NDs by 2--3 MHz (Fig.~\ref{fig2}\textbf{f}), indicating reduced crystal strain in the \ce{^12C}, N-NDs.
The effect of the geomagnetic field on \textit{E} was insignificant due to the random NV quantization axes (see Supplementary Information).
Unexpectedly, the mean \textit{D} of the \ce{^12C}, N-NDs was 1--2 MHz higher than those of the type-Ib NDs (Fig.~\ref{fig2}\textbf{g}). 
Because the mean \textit{D} of Ib-600 was higher than that of Ib-100 as well, this increment in the mean \textit{D} possibly results from the ND size difference among the \ce{^12C}, N-and type-Ib NDs; however, the absolute value of \textit{D} has not been studied in the context of ND size, and the exact origin of the increased \textit{D} continues to remains unclear.

\begin{figure}[th!]
    \centering
    \includegraphics[width=8.5 cm]{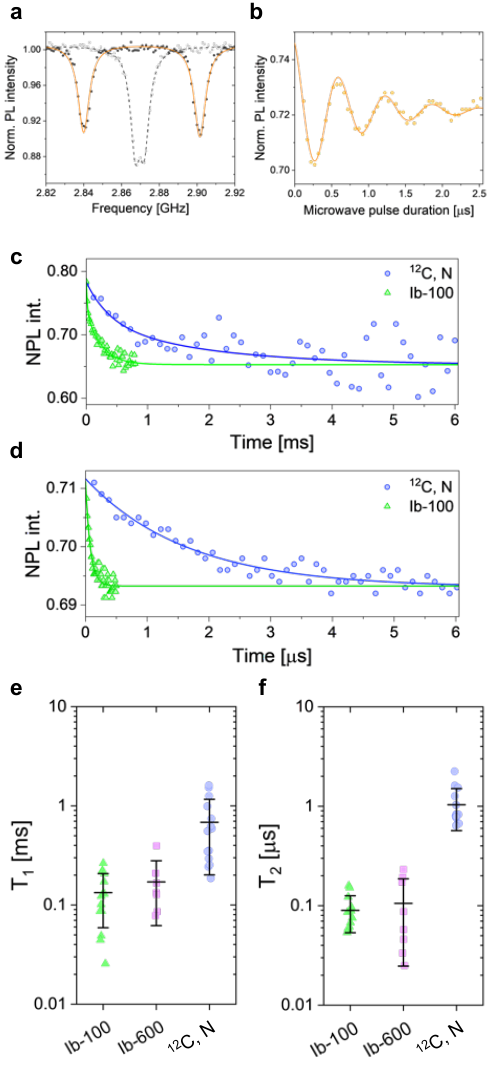}
    \caption{
    (a) ODMR spectra and double-Lorentzian fits of a representative \ce{^12C}, N-ND with (black dots and orange solid line) and without (gray dots and gray dashed line) an external magnetic field. 
    (b) A typical Rabi oscillation observed in the \ce{^12C}, N-NDs. 
    A solid line is a sine-damp fit. 
    Representative profiles of (c) all-optical $T_1$ relaxometry and (d) $\pi/2$-spin echo measurements for the \ce{^12C}, N- and Ib-100 NDs. 
    Statistical plots of $T_1$ (e) and $T_2$ (f) relaxation times for the \ce{^12C}, N-, Ib-100 and Ib-600 NDs.
    }
    \label{fig3}
\end{figure}

To further evaluate the NV spin properties of the \ce{^12C}, N-NDs, we performed pulsed ODMR experiments and determined the longitudinal ($T_1$) and transverse ($T_2$) relaxation times of the \ce{^12C}, N-and Ib-100 NDs (Figs.~\ref{fig3}\textbf{a--d}).
An external magnetic field was applied in a controlled orientation to split the spectra into two peaks (Fig.~\ref{fig3}\textbf{a}).
By addressing the lower-energy peak, we measured Rabi nutation to determine the duration of $\pi$-pulse used in the subsequent $T_1$ and $T_2$ measurements, where the typical duration of the $\pi$-pulse is 300--400 ns. (Fig.~\ref{fig3}\textbf{b}).
Figures~\ref{fig3}\textbf{c, d} show the representative $T_1$ and $T_2$ profiles for the \ce{^12C}, N-and Ib-100. 
Both profiles exhibit a substantial extension of the relaxation times in the \ce{^12C}, N-NDs.
$T_1$ profiles were fit with a biexponential decay, and $T_2$ profiles with a stretched exponential decay (see Methods). 
We performed statistical analysis by measuring a number of NDs as shown in Figs.~\ref{fig3}\textbf{e, f}.
The \ce{^12C}, N-NDs exhibit $T_1^{\rm max}$ = 1.6 ms and $T_2^{\rm max}$ = 2.7 $\si{\us}$ in maximum, which are close to the bulk-limited relaxation times for the present nitrogen concentration of 30--60 ppm ($T_1 \approx 3 \ \si{\ms}$~\cite{jarmola2012temperature}, $T_2 \approx 3$--$5$ \si{\us}~\cite{bauch2020decoherence}.). 
$T_1$ of the \ce{^12C}, N-NDs ($0.68 \pm 0.48$ ms) is more than 4--5 times longer than those of the type-Ib NDs (Ib-100: $0.13 \pm 0.07$ ms, Ib-600: $0.17 \pm 0.11$ ms) , and $T_2$ of the \ce{^12C}, N-NDs ($1.6 \pm 0.6 \ \si{\us}$) is more than 11 times longer than those of the type-Ib NDs (Ib-100: $0.14 \pm 0.05 \ \si{\us}$, Ib-600: $0.11 \pm 0.08 \ \si{\us}$). 
These values represent a remarkable improvement in the spin relaxation times at [NV] = 0.6--1.3 ppm.

\begin{figure*}[th!]
    \centering
    \includegraphics[width=17 cm]{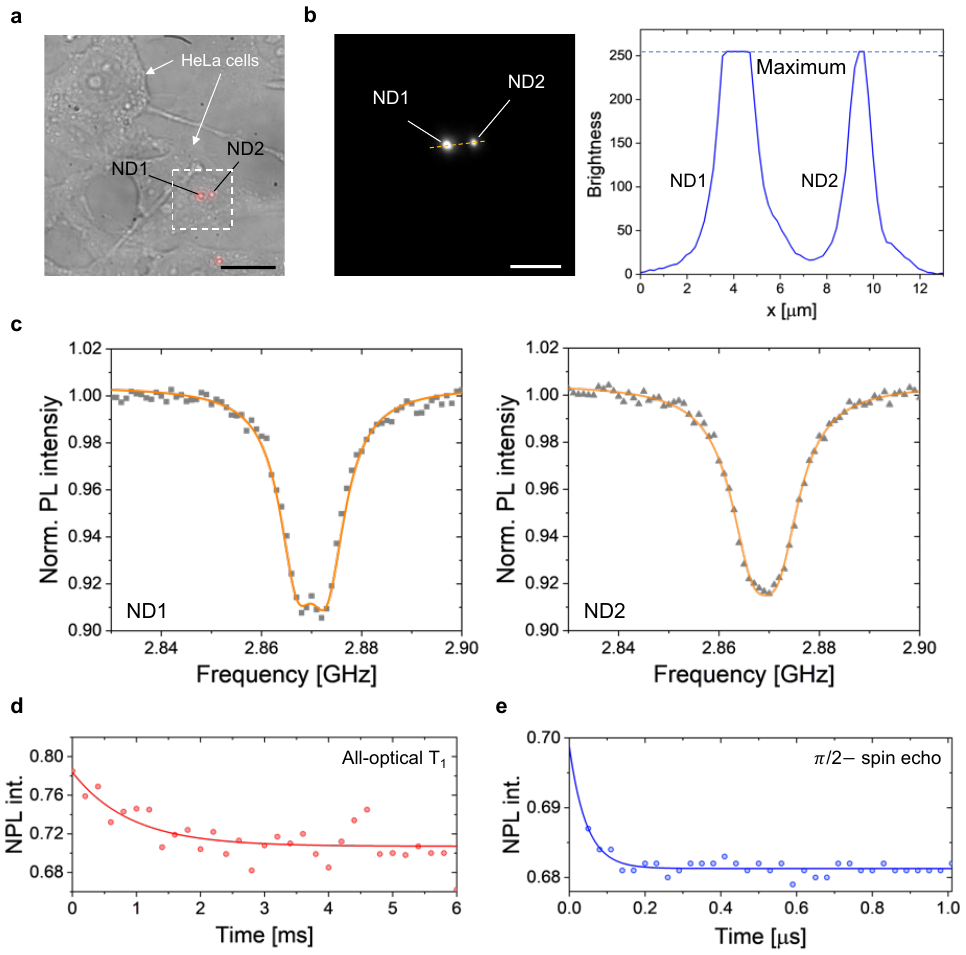}
    \caption{
    (a) Merged microscope image of gray bright-field and red fluorescence for HeLa cells uptaking \ce{^12C}, N-NDs (Scale bar: 25\,\si{\um}). 
    (b) Red fluorescence (ND1, ND2) in the white dotted box of Fig.~\ref{fig4}(b) (Scale bar: 10\,\si{\um}), and a cross-section along the yellow dotted line exhibiting the brightness of ND1 and ND2 (maximum value: 256).
    (c) In-situ ODMR spectra of ND1 and ND2 in the live cells.
    (d) All-optical $T_1$ relaxometry and (e) $\pi /2$-spin echo profiles for ND2.
    }
    \label{fig4}
\end{figure*}

Having characterized the optical-spin properties of the \ce{^12C}, N-NDs, we feed these NDs to cultured HeLa cells to perform CW and pulsed ODMR experiments inside live cells.
We fabricated a culture device by bonding a multiwell acrylic frame to coverslips with a notch-shaped antenna and cultured the HeLa cells in the wells (see Fig.~\ref{Cell culture protocol and other results}\textbf{a}).
Subsequently, we introduced \ce{^12C}, N-NDs into the cells via endocytosis (see Methods). 
Figure~\ref{fig4}\textbf{a} shows a merged bright-field and red-fluorescence image of ND-labeled HeLa cells. 
The two NDs in Fig.~\ref{fig4}\textbf{a} (designated as ND1 and ND2) are sufficiently bright to distinguish in the presence of autofluorescence in the cells.
Figure~\ref{fig4}\textbf{b} indicates that the brightness of ND1 and ND2 on the yellow dotted lines in Fig.~\ref{fig4}\textbf{a} is saturated or near saturation, whereas the background fluorescence is nearly zero. 

We measured and observed narrow and deep ODMR spectra of ND1 and ND2 inside the cells (Fig.~\ref{fig4}\textbf{c}).
Fluorescence fluctuations owing to Brownian motion were observed during the CW-ODMR measurements. 
We performed pulsed ODMR measurements on ND2 under an external magnetic field in a similar manner to that described above.
The Brownian motion significantly affected the pulsed ODMR data when the microwave was applied because the random fluctuation of the ND orientation appears as dephasing during the signal integration ~\cite{maclaurin2013nanoscale} (see Figs.~\ref{Cell culture protocol and other results}\textbf{c--e}).
All-optical $T_1$ relaxometry is an exception, which utilizes time-dependent spin polarization after optical initialization~\cite{sigaeva2022diamond} (Fig.~\ref{fig4}\textbf{d}). 
The profile exhibited a millisecond-scale relaxation.
Double exponential fitting gives a value of $T_1 = 0.87$ ms, which is six times larger than the mean $T_1$ value of type-Ib NDs.
In contrast, the $T_2$ relaxation profiles exhibit a substantial shortening of $T_2$ owing to the Brownian motion. 
Figure~\ref{fig4}\textbf{e} shows the relaxation profile of the $\pi/2$-spin-echo sequence. 
Single-exponential fitting yields $T_2 = 67.5$ ns, which is one order of magnitude shorter than the mean $T_2$ value determined above. 
The difficulties associated with the Brownian motion of NDs during pulsed ODMR measurements have been previously discussed~\cite{kayci2014electron}; however, to date, this phenomenon has not been actively investigated owing to the very short $T_2$ in type-Ib NDs. 
The NV spin coherence of the present \ce{^12C}, N-NDs will enable to develop advanced pulsed ODMR quantum-sensing protocols against the Brownian motion.

\section{Discussion}
\label{sec4}
This study demonstrates a remarkable improvement in the NV spin properties of ultrabright NDs with [NV] = 0.6--1.3 ppm.
Microwave irradiation may result in heat generation and damage biological samples. Thus, the observed high contrast of the CW-ODMR spectra and up to a 16-fold reduction in microwave power are promising features for biological applications.
The \ce{^12C}, N-NDs exhibit long average spin-relaxation times of $T_1 = 0.68$ ms and $T_2 = 1.6 \ \si{\us}$, which are close to the bulk diamond limits and are longer than those of the conventional type-Ib NDs by factors of 6 and 11, respectively.
This enhancement of the NV spin coherence is principally attributed to the spin-impurity control. 
We also confirm that $T_2$ of the Ib-100 and Ib-600 NDs are close to the bulk-limited values calculated from their [NV] = 300--540 ppm, which is estimated from their infrared absorption spectra (see Supplementary Information).
The spin-impurity control is thus critical to improve the NV spin characteristics of the NDs with a size of $\gtrsim 80$ nm. 
The present $T_1$ extension increases the relaxometric signal-to-noise ratio (SNR) in Gd$^+$ detection~\cite{steinert2013magnetic}. Further, the \ce{^12C}, N-NDs can improve the SNR by a factor of six compared with that of the type-Ib NDs, thus potentially lowering the detection limit for reactive oxygen species~\cite{sigaeva2022diamond}.
Similarly, the 10-fold $T_2$ extension increases the measurement sensitivity in AC magnetometry by a factor of three \cite{tetienne2018spin}, thus allowing the implementation of various $T_2$-based quantum measurement protocols, such as advanced spin-echo and thermal-echo quantum sequences in NDs~\cite{knowles2016demonstration, abobeih2019atomic}.

The proposed \ce{^12C}, N-NDs pose some technical challenges, including: (i) the inhomogeneity in the NV characteristics should be further minimized, (ii) pulsed-ODMR protocols against Brownian motion must be established, and (iii) large-scale production of NDs is required.
First, the inhomogeneity in the NV characteristics may arise from the spatial variations in the NV spins inside the original single crystalline diamonds because of several factors, such as different growth sectors, atomic concentrations, and dislocations~\cite{pham2012enhanced, capelli2019increased}.
Further, surface spin noises can fluctuate the NV characteristics of each ND because inhomogeneous surface termination, e.g., different types and contents of dangling bonds, causes variations in the surface spin noise~\cite{rosskopf2014investigation, zheng2022coherence}; these inhomogeneity factors may vary $T_1$ and $T_2$ (Figs.~\ref{fig3}\textbf{e, f}).
In the present study, the variations in $T_1$ and $T_2$ observed in the case of the \ce{^12C}, N-NDs are comparable with those for the Ib-100 and Ib-600 NDs. Therefore, these variations need to be further minimized to improve the measurement accuracy of quantum sensors employed in biomedical applications~\cite{sigaeva2022diamond}.

Inhomogeneity in the NV spin environment also affects the pulsed ODMR, particularly through $T_{2}^*$. 
According to previous studies, $T_{2}^*$ of the NV ensembles in high-NV-density type-Ib NDs was reported to be extremely short ($< 100$ ns) based on the assumption that it could not be measured ~\cite{liu2019coherent}. 
Further, the measurement of pulsed ODMR, including Rabi oscillations and $3\pi /2$-spin echo, becomes challenging when the ND size increases (Ib-100 and Ib-600; see Figs.~\ref{Ib spin echo fitting}\textbf{a--d}, Fig.~\ref{Ramsey}\textbf{a}) because of the increased inhomogeneity in the NV spin environment. 
In contrast, we observed clear Rabi oscillations and spin-echo inversion of the $3\pi /2$-spin echo in the \ce{^12C}, N-NDs (Figs.~\ref{12C and Ib fitting}\textbf{j--i}, Fig.~\ref{Ramsey}\textbf{a}), indicating long $T_{2}^*$ for the NV ensembles.
Indeed, we observed Ramsey oscillations for a certain \ce{^12C}, N-ND with $T_{2}^* = 140$ ns (Figs.~\ref{Ramsey}\textbf{b, c}), while a statistical analysis was impossible because of the low success probability; this particular ND might have an excellent NV spin environment owing to the present spin-impurity control.

Second, pulsed ODMR measurements must be optimized for NDs exhibiting Brownian motion. 
In the pulsed ODMR experiments, both the $T_1$ and $T_2$ profiles show substantial shortening of the relaxation times, except for the all-optical $T_1$ relaxometry, that is, when microwave spin manipulations are involved.
NDs exhibit Brownian motion inside the cells, and the rotation of the NV orientation relative to the microwave magnetic field fluctuates. 
During signal integration in the present measurement protocol, this effect modulates the spin signal, effectively shortening the microwave $T_1$ and spin-echo decays~\cite{maclaurin2013nanoscale}. Only all-optical $T_1$ relaxometry provides a profile similar to that observed for stationary NDs.
Therefore, integrating pulsed ODMR protocols is necessary to adapt to random orientations and motion ~\cite{konzelmann2018robust,schwartz2018robust}.

Finally, the large-scale production of \ce{^12C}, N-NDs is required for further biochemical applications.
Batch reactions for the surface functionalization of NDs require more than 10 mg of NDs to obtain higher yields ~\cite{komatsu2023poly}. 
Currently, the number of single crystals with reduced spin impurities and high-density NV is insufficient because of the special production processes involving \ce{^12C} methane and the absence of a mass production line.
Furthermore, the milled NDs are centrifuged to separate the fractions with different sizes of NDs, which substantially reduces the final amount of NDs. 
The optimization of both the production process and the cost is important for achieving large-scale production.

In conclusion, we demonstrated the development of ultrabright spin-impurity controlled NDs containing $[\rm  NV]_{\ce{^12C}, N}$ = 0.6--1.3 ppm with spin impurities of [\ce{^12C}] = 99.99\% and [N] = 30--60\,ppm.
The NDs exhibit a 50--700 nm size range and are readily used for the fluorescent labeling of cultured cells.
We demonstrated a remarkable improvement of the NV spin quality as compared to that of the conventional type-Ib NDs including narrow and deep CW-ODMR spectra and the extended average spin-relaxation times of $T_1 = 0.68\,\si{\ms}$ and $T_2 = 1.6\,\si{\us}$ (1.6 ms and 2.7 \si{\us} in maximum).
We observed that these NDs used 16 times less microwave power to reach a 3\% ODMR contrast than their type-Ib counterparts.
Using these \ce{^12C}, N-NDs, we performed ODMR measurements (CW-ODMR, $T_1$ and $T_2$) inside the cells.
These results successfully demonstrate the pertinence of quantum-grade NV spin properties for quantum sensing in potential biological applications.

\clearpage
\section{Methods}
\label{sec6}
\subsection{ND preparation}
${\rm ^{12}C}$-isotope-enriched single-crystalline bulk diamonds with controlled nitrogen concentrations were synthesized using the HPHT method as described previously~\cite{ikeda2016optical}, with minor modifications for single-crystal growth.
The ${\rm ^{12}C}$ enrichment was 99.99\% and the nitrogen concentration was 30--60 ppm, as confirmed via secondary ion mass spectroscopy (SIMS) after HPHT growth.
NV centers were generated in these bulk diamonds using electron beam irradiation (3 MeV, $1 \times 10^{18}\,\mathrm{cm^{-18}}$) under ambient conditions, followed by vacuum annealing at 900 $\si{\degreeCelsius}$ for 1 h~\cite{deguchi2023compact}.
These bulk diamonds were pulverized, followed by suspension in water to obtain \ce{^12C}, N-NDs. Type-Ib NDs with mean sizes of 100 and 600\,nm were purchased from Adámas Nanotech. (NDNV100nmHi, NDNV600nmHi).

\subsection{ND coating on the coverslips and AFM}
To determine the NV concentration of \ce{^12C}, N-NDs, coverslips with engraved island grids on one side was used.
A small droplet of the ND suspension was drop-casted onto a coverslip, enabling the utilization of the same NDs for brightness confirmations and AFM measurements.
The topographies of the spin-coated samples were determined using AFM in the tapping mode (SPA400, Hitachi High-Tech Corporation) after the ODMR measurements.
The height of the ND topography was regarded as the ND size to avoid the tip convolution effect~\cite{canet2014correction, stehlik2015size, tsukahara2019removing}.
The images were collected at a scan rate of 0.1--0.5 Hz.

\subsection{Optical and ODMR measurements}
The optical properties and ODMR of the NDs were measured using a lab-built confocal fluorescence microscope with a microwave excitation system based on previous studies ~\cite{fujiwara2020real, fujiwara2020real_2, oshimi2022glass}.
A 532-nm laser was used to excite the NDs with an intensity in the 5--10 $\si{kW.cm^{-2}}$ range, which corresponds to an optical saturation parameter of $s = 0.05-0.10$~\cite{oshimi2022glass,dreau2011avoiding}.
For the excitation and the fluorescence collection, a 50x dry objective with a numerical aperture of 0.7 mounted on a piezo actuator (Piezosystemjena, MIPOS 100) for fine z-axis adjustment was used.
A voice-coil-driven fast-steering mirror (Optics In Motion, OIM101) was used for fast xy-scanning of the laser. 
NV fluorescence was filtered using a dichroic beam splitter (Semrock, FF560-FDi01) and a long-pass filter (Semrock, BLP01-635R-25) to remove residual green laser scattering.
The fluorescence was coupled to an optical fiber (Thorlabs, 1550HP) and detected using a single-photon counting module (Excelitas, SPCM-AQRH-14), and its spectra were measured using a spectrometer equipped with a charge-coupled device camera (Princeton Instruments, PIX256OE-SF-Q-F-A).
The output from the photon-counting module was fed into a board system (National Instruments USB-6343 BNC).
The laser scanning was controlled using a lab-built program using GPScan~\cite{ferrand2015gpscan} in part. 
In the ODMR measurements for both the continuous-wave (CW) and pulsed modes, microwaves were generated using a signal generator (Rohde \& Schwarz, SMB100A) and sent to radiofrequency (RF) switches (Mini-circuit, ZYSWA-2-50DRS and General Microwave, F9160) triggered by a bit-pattern generator (SpinCore, PBESR-PRO-300). 
The signal was then amplified using a 45 dB amplifier (Mini-circuit, ZHL-16W-43+).
In the CW mode, microwave excitation was gated using the RF switches to suppress noise (200\,\si{\us} for microwaves on and off). 
In the pulsed mode, external magnetic fields of 5--15 G were applied along the NV quantization axis using a small neodymium magnet mounted on a multi-axis manual stage to lift the degeneracy of the magnetic sublevels and detect Rabi-nutations in the pulsed measurements.
The pulse operation of the excitation laser was implemented using an acousto-optic modulator (AOM, G\&H, 3200-121), and NV spins were optically initialized using a 3 \si{\us} pulse width.
All pulse sequences are presented in Fig.~\ref{Optical setup and sequence}\textbf{b--d}. 
To determine the duration of $\pi$ pulse for the NV spins, Rabi measurements were performed, and the profile was fitted with a sine-damping function. 
$T_{2}$ was measured using spin-echo sequences, and $\pi/2 - \pi - \pi/2$ ($\pi/2$-spin echo) and $\pi/2 - \pi - 3\pi/2$ ($3\pi/2$-spin echo) sequences were measured to cancel common-mode noise~\cite{alsid2019photoluminescence}.
$T_{1}$ was measured using spin-polarization relaxometry sequences. We measured all-optical $T_1$ and microwave $T_1$ relaxometry sequences, separately and cancelled the common-mode noise by subtraction~\cite{jarmola2012temperature}.

\subsection{CW-ODMR spectral analysis}
The observed ODMR spectra were fitted using a double-Lorentzian function ($y (x)$) composed of two Lorentzian functions, $L_{1,2} (x)$: $y (x) = y_{0} +L_1 (x) +L_2 (x)$, and $L_{1,2} (x) = (2A_{1,2}/\pi)[w_{1}/(4(x-x_{1,2})^2+w_{1,2}^2)]$, where, $y_0$, $A_{1,2}$, $w_{1,2}$, and $x_{1,2}$ are the offset, peak area, linewidth, and peak position frequency of $L_{1,2} (x)$, respectively.
The following boundary conditions were applied to the fitting: $2.862 \leq x_{1} \leq 2.870$ and $ 2.870 \leq x_{2} \leq 2.878$. 
The ``Norm.~PL~intensity'' in Fig.~\ref{fig2}\textbf{d} were determined by taking the mean of the two peaks $[y(x_1) + y(x_2)]/2$.
In this study, the spectral parameters for the peak splitting \textit{E} and zero-field splitting \textit{D} were determined using the fitting parameters $\textit{E} = x_{2}-x_{1}$ and $\textit{D} = (x_{1} + x_{2})/2$, respectively (Figs.~\ref{fig2}\textbf{e, f}).
$E$ is the strain-induced parameter~\cite{teissier2014strain, doherty2014electronic,sotoma2018enrichment}.
Further details of the fitting procedure are provided in the Supplementary Information (Fig.~\ref{Fitting procedure}\textbf{a}).

\subsection{Analysis method for the $T_1$ and $T_2$}
The raw $T_1$ relaxation profiles showed large amount of noise, which destabilized the subsequent fitting processes.
These noises were numerically filtered by taking a moving average over nine data points (out of a total of 50 points), and the filtered profiles measured with the all-optical $T_1$ and microwave $T_1$ sequences were subtracted (see Fig.~\ref{12C and Ib fitting}\textbf{a--f}). 
The subtracted $T_1$ profiles were fitted using a two-phase exponential decay, $C = C_{0} + A_{1}\exp{(-(\tau-t_{0})/t_{1})} + A_{2}\exp{(-(\tau-t_{0})/t_{2}})$, where $C_0$ is the common offset and $A_{1,2}$,and $t_{1,2}$ are amplitudes, and time constants, respectively.
Out of the two decay times, the longer one was used as $T_1$ ($T_1 = \max(t_{1}, t_{2})$) as previously described~\cite{robledo2011spin, ziem2013highly, perona2020nanodiamond, barbosa2023impact, barbosa2023temperature}.
Similarly, the $\pi/2$- and $3\pi/2$-spin-echo profiles were filtered by moving the average over five data points (out of a total of 50 points) before their subtraction.
For the \ce{^12C}, N-NDs, the subtracted profiles were fitted using the stretched exponential decay $\exp{((-\tau/T_{2})^{1.5})}$ to determine $T_2$~\cite{bauch2020decoherence} (see Figs.~\ref{12C and Ib fitting}\textbf{a--c}).
For the $T_2$ determination of the type-Ib NDs, this fitting is only applied to  the $\pi/2$-spin echo profiles because $3\pi/2$-spin echo sequence did not provide spin-echo amplitudes owing to the short $T_{2}^*$ (see Supplementary Information).

\subsection{ND labeling of HeLa cells}
HeLa cells were cultured in a cell culture medium (DMEM containing 4500 mg/mL glucose, 10\% FBS, 100 U $\si{\mL}^{-1}$ penicillin/streptomycin, and phenol red) in a well that was fabricated on a coverslip, with a notch-shaped antenna on the other side (see Fig.~\ref{Cell culture protocol and other results}\textbf{a}).
The glass surface of the device (well side) was coated with collagen to improve cell adhesion. 
ND labeling of cells was performed using the method described previously~\cite{nishimura2021wide, yukawa2020quantum}.
A suspension of \ce{^12C}, N-NDs was added to the culture medium, and the cells were incubated at 37 $\si{\degreeCelsius}$ and 5\% $\mathrm{CO_2}$ for 24 h.
The cells were then washed gently three times with phosphate-buffered saline and immersed in a culture medium without phenol red to perform the ODMR measurements.

\subsection{FTIR measurments}
The nitrogen concentration in the type-Ib NDs was estimated by measuring 150-\si{\um}-sized type-Ib diamond micro-crystals (Adámas Nanotech., MDNV150umHi) from the same product line using the Fourier transform infrared (FTIR) spectroscopy, as previously reported~\cite{chertkova2015situ, sano2019evolution}.
A Jasco FTIR6200-IRT7000 micro Fourier-transform spectrometer was used with a KBr/Ge beam splitter, a ceramic light source, and MCT detector. 
The micro-crystals were placed on a copper mesh with pore sizes of $85\,\si{\um}$, and near-infrared (NIR) light was focused with a spot size of 50--100 \si{\um} through Cassegrain optics in the transmission mode.
The sample chamber was purged with nitrogen gas to minimize atmospheric background in the spectra.
The spectral data were accumulated 512 times with a spectral resolution of $4\,\si{cm^{-1}}$.
From the obtained spectrum, the nitrogen concentration [N] (ppm) was determined using the following relationship: $(\mu_{1130 \si{cm^{-1}}}/\mu_{2120 \si{cm^{-1}}}) \times 5.5 \times 25$, where $\mu$ is absorption intensity for the given wavenumber~\cite{liang2005synthesis}.

\section*{Acknowledgments}
The authors thank 
I. C. Barbosa, O. Benson, H. Ishiwata, Y. Masuyama, R. Schirhagl, S. Sakakihara, T. Schröder, T. Teraji, S. Yamashita for fruitful discussions and technical assistance with the experiments.
A part of this work was supported by ARIM of MEXT (JPMXP1223OS1022) and IPM, Okayama University under joint-use facilities framework.
This study was partially supported by JSPS-KAKENHI (M.F. and Y.S. 20H00335. M.F.: 20KK0317. Y.S.: 21H05599.) and JST-ASPIRE (Y.S. and M.F.: JPMJAP2339).
K.O. acknowledges funding from JST (JPMJFS2128) and JSPS (23KJ1607). 
S.M. acknowledges funding from JST (JPMJSP2126).
M.F. acknowledges funding from NEDO (JPNP20004), AMED (JP23zf0127004), JST (JPMJMI21G1), RSK Sanyo Foundation, and Asahi Glass Foundation.
Y.S. acknowledges the funding from JST-PRESTO (JPMJPR20M4).

\section*{Author contributions}
K.O., M.T., Y.N., and M.F. conceptualized the research.
M.T., H.T., and Y.N. prepared \ce{^12C}, N-NDs.
K.O., H.N., S.M., H.K., and M.F. performed NV spin characterization and cellular experiments. 
K.O. and T.A. performed the ND size determination using AFM. 
K.O., Y.S., and M.F. analyzed the NV spin data.
All the authors participated in the discussion and writing of the manuscript.

\clearpage
\appendix
\renewcommand{\thefigure}{S\arabic{figure}}
\renewcommand{\thetable}{S\arabic{table}}
\renewcommand{\theequation}{S\arabic{equation}}
\setcounter{figure}{0}  
\setcounter{table}{0}  
\setcounter{equation}{0}  

\onecolumngrid
\section*{Supplementary Information}

\section{Determination of the NV concentration in NDs with size--brightness correlation}
\begin{figure*}[h!]
    \centering
    \includegraphics[width=14cm]{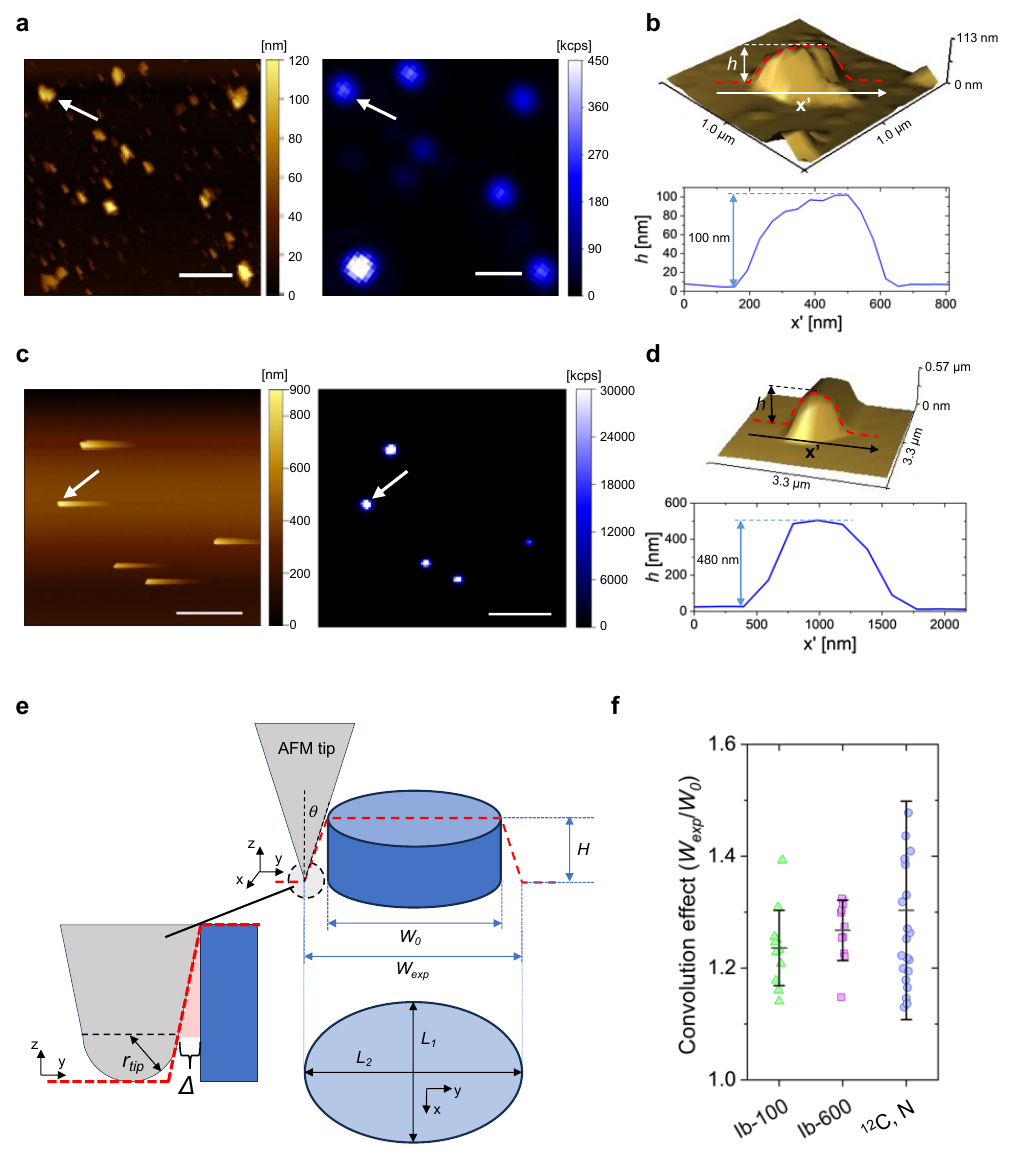}
    \caption{
    (a) AFM topography (left) and the corresponding confocal fluorescence (right) images of Ib-100. Scale bar; 1\,\si{\um}. 
    (b) Three-dimensional visualization of the topography of the ND indicated by the white arrow in Fig.~\ref{AFM image for Ib}\textbf{a} (top) with a cross-section along $x'$ axis (bottom). 
    (c) AFM topography (left) and the corresponding confocal fluorescence (right) images of Ib-600. Scale bar; 10\,\si{\um}. 
    (d) Three-dimensional visualization of the topography of the ND indicated by the white arrow in Fig.~\ref{AFM image for Ib}\textbf{c} (top) with a cross-section along $x'$ axis (bottom).
    (e) Schematic of the AFM measurements with the convolution effect.
    $W_{exp}$, $W_{0}$, $H$, $r_{tip}$, and $\theta$ are the experimental, true length in the xy-imaging plane of the measured sample, height of the sample from the xy-imaging plane, AFM tip radius, and tip angle, respectively.
    $L_1$ and $L_2$ are the ND width in the x and y axes, respectively.
    $r_{tip} = 10\,\si{nm}$, $\theta = 20^{\circ}$.
    Red dotted line: the convolution effect of the AFM tip.
    (f) Statistical plots of the convolution effect ($W_{exp}/W_{0}$) for Ib-100, Ib-600 and \ce{^12C}, N-NDs.
    Mean and standard deviation (1$\sigma$) are indicated in the statistical plots.
    }
    \label{AFM image for Ib}
\end{figure*}

Figures~\ref{AFM image for Ib}\textbf{a--d} show AFM topology and confocal images of the type-Ib NDs (Ib-100, Ib-600).
These AFM topography images, including Figs.~\ref{fig1}\textbf{d, e} for \ce{^12C}, N-NDs, have a common artifact: convolution caused by interaction between sample and AFM tip.
Figure~\ref{AFM image for Ib}\textbf{e} shows a schematic of this convolution effect in the AFM measurements of NDs. 
The experimentally determined width of the NDs ($W_{exp}$) in the xy-imaging plane can be written as~\cite{canet2014correction}: 
\begin{equation}
\label{convolution}
    \begin{aligned}
    \frac{1}{2} W_{exp} &= r_{tip} + \Delta + \frac{1}{2} W_{0}\\
    \Delta &= (H - r_{tip})\tan\theta
    \end{aligned}
\end{equation}
where $W_{0}$, $H$, $r_{tip}$, and $\theta$ are ND width in the xy-imaging plane, ND's height from the xy plane (glass surface), tip radius, and tip angle, respectively.
In the real NDs, the widths in the x and y axes ($L_{1,2}$) differ because of the asymmetric shape and we adopted their mean values as $W_{exp}$ for simplicity. 
The term of $r_{tip} + \Delta$ indicates the tip convolution.
From this analysis, we calculated $W_{exp} / W_{0}$ to show how much convolution effect exists in the present three types of NDs (Ib-100, Ib-600, \ce{^12C}, N-NDs). 
As shown in Fig.~\ref{AFM image for Ib}\textbf{f}, the results indicate that the convolution effect is comparable to each other among these samples.

\begin{figure*}[th!]
    \centering
    \includegraphics[width=17cm]{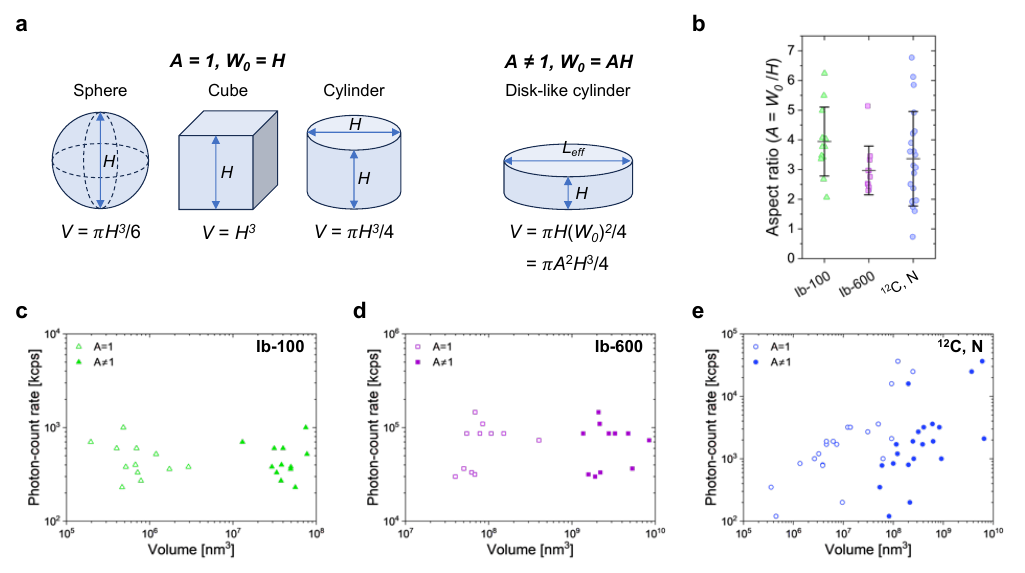}
    \caption{
    (a) Schematic images of the typical ND morphology with their parameters defining the shapes. 
    $A$, $V$, $H$, and $W_{0}$ indicate aspect ratio, volume, width, and height, respectively.
    (b) Statistical plots of the aspect ratio for Ib-100, Ib-600 and \ce{^12C}, N-NDs.
    Mean and standard deviation (1$\sigma$) are indicated in the statistical plots.
    Plots of the ND photon counts against the ND volume in the case of the cylinder with $A = 1$ and $A \neq 1$ for (c) Ib-100, (d) Ib-600, (e) \ce{^12C}, N-NDs.
    }
    \label{Various structure for NDs with aspect ratio}
\end{figure*}

After determining $W_{0}$, we calculated [NV] as follows. 
We first assumed that the fluorescence intensity is proportional to the number of NVs as the optical excitation intensity is far below the saturation ($s \approx 0.05$). 
As the number of NVs is proportional to the ND volume, the fluorescence intensity can be written as 
\begin{equation}
    I \propto [NV] V, 
\end{equation}
where $I$ and $V$ are the fluorescence photon-count rate and the volume of NDs, respectively.
By taking a ratio of the fluorescence intensity for the two types of the NDs, [NV] can be determined as follows: 
\begin{equation}
\label{V and photon relation}
    [NV]_j = [NV]_i \frac{I_j}{I_i} \frac{V_i}{V_j},  
\end{equation}
where the subscripts $i, j$ indicate the type of the NDs.
By introducing aspect ratio ($A = W_{0}/H$), $V$ can be expressed as
\begin{equation}
\label{volume with aspect ratio}
    \begin{aligned}
    V = k H^3 g(A)
    \end{aligned}
\end{equation}
where $k$ and $g(A)$ denote morphology dependent factor and $A$-dependent conversion factor, respectively (e.g, $k = \pi/6$, $g(A) = 1$ for sphere, $k = 1$, $g(A) = 1$ for cube, $k = \pi/4$, $g(A) = 1$ for cylinder, as shown in Fig.~\ref{Various structure for NDs with aspect ratio}\textbf{a}).

We then estimated the NV concentration of the \ce{^12C}, N-NDs. 
In Fig.~\ref{fig1}\textbf{g}, the fluorescence photon-count rate of Ib-600 NDs was approximately $150$ times more than that of Ib-100 ($I_{\rm Ib600}/I_{\rm Ib100} \approx 150$).
Provided both of the ND morphology as $A=1$ type such as cubic, Ib-600 NDs were found to be approximately $130$ times larger in volume than Ib-100 NDs ($V_{\rm Ib600}/V_{\rm Ib100} \approx 130$).
Through Eq.~\ref{V and photon relation} and Eq.~\ref{volume with aspect ratio}, these parameters give $[NV]_{\rm Ib600} \approx 1.2 [NV]_{\rm Ib100}$, comparable to the ratio determined from their specification sheet ($[\rm NV]_{\rm Ib100} = 3\,ppm$, $[\rm NV]_{\rm Ib600} = 3.5\,ppm$).
Similarly, assuming the premise of \ce{^12C}, N-NDs having $A=1$ type morphology, we estimated $[\rm NV]_{\ce{^12C}, N}$ to be $0.26 
[\rm NV]_{\rm Ib100} \approx 0.6$ ppm or $0.20 [\rm NV]_{\rm Ib100} \approx 0.8$ ppm by taking a reference to Ib-100 or Ib-600 NDs, respectively.

We also estimated $[\rm NV]_{\ce{^12C}, N}$ for a $A \neq 1$ morphology such as disk-like ND structures ($k = 1$, $g(A) = A^2$ in Eq.~\ref{volume with aspect ratio}).
Figure~\ref{AFM image for Ib}\textbf{b} shows statistical plots of $A$ by analyzing the AFM data of the NDs. 
From this data, we obtained $A^{mean}$ for Ib-100, Ib-600 and \ce{^12C}, N-NDs to be $3.9 \pm 1.2$, $2.9 \pm 0.8$ and $3.3 \pm 1.6$, respectively.
These values are comparable to the previous study about 140 nm NDs, where the NDs have the disk-like shape with $A \approx 3$~\cite{eldemrdash2023fluorescent}.
Note that we analysed only one length of NDs (either of $L_{1}$ and $L_{2}$) using the tailed AFM topography image caused by the measurement limitation of AFM tapping mode (Fig.~\ref{AFM image for Ib}\textbf{c}), so that it is possible that the calculated aspect ratio of Ib-600 can become slightly smaller value than that of Ib-100.
Figures~\ref{Various structure for NDs with aspect ratio}\textbf{c--e} show volume-brightness correlation plots of the type-Ib NDs and \ce{^12C}, N-NDs for $A = 1$ and $A \neq 1$ (disk-like cylinder) cases.
The scatter plots were entirely shifted to the lager side for all of the ND types, when the volume estimation method was switched from $A = 1$ to $A \neq 1$.
Using this disk-like-cylinder shape with $A \neq 1$, we obtained $0.44 [\rm NV]_{\rm Ib100} \approx 1.3$ ppm or $0.21 [\rm NV]_{\rm Ib600} \approx 0.6$ ppm by taking a reference to Ib-100 or Ib-600 NDs, respectively.
Therefore, we determined [NV] of the present \ce{^12C}, N-NDs was 0.6--1.3 ppm.

\clearpage
\section{ODMR Experimental Setup}
\begin{figure*}[h!]
    \centering
    \includegraphics[width=14cm]{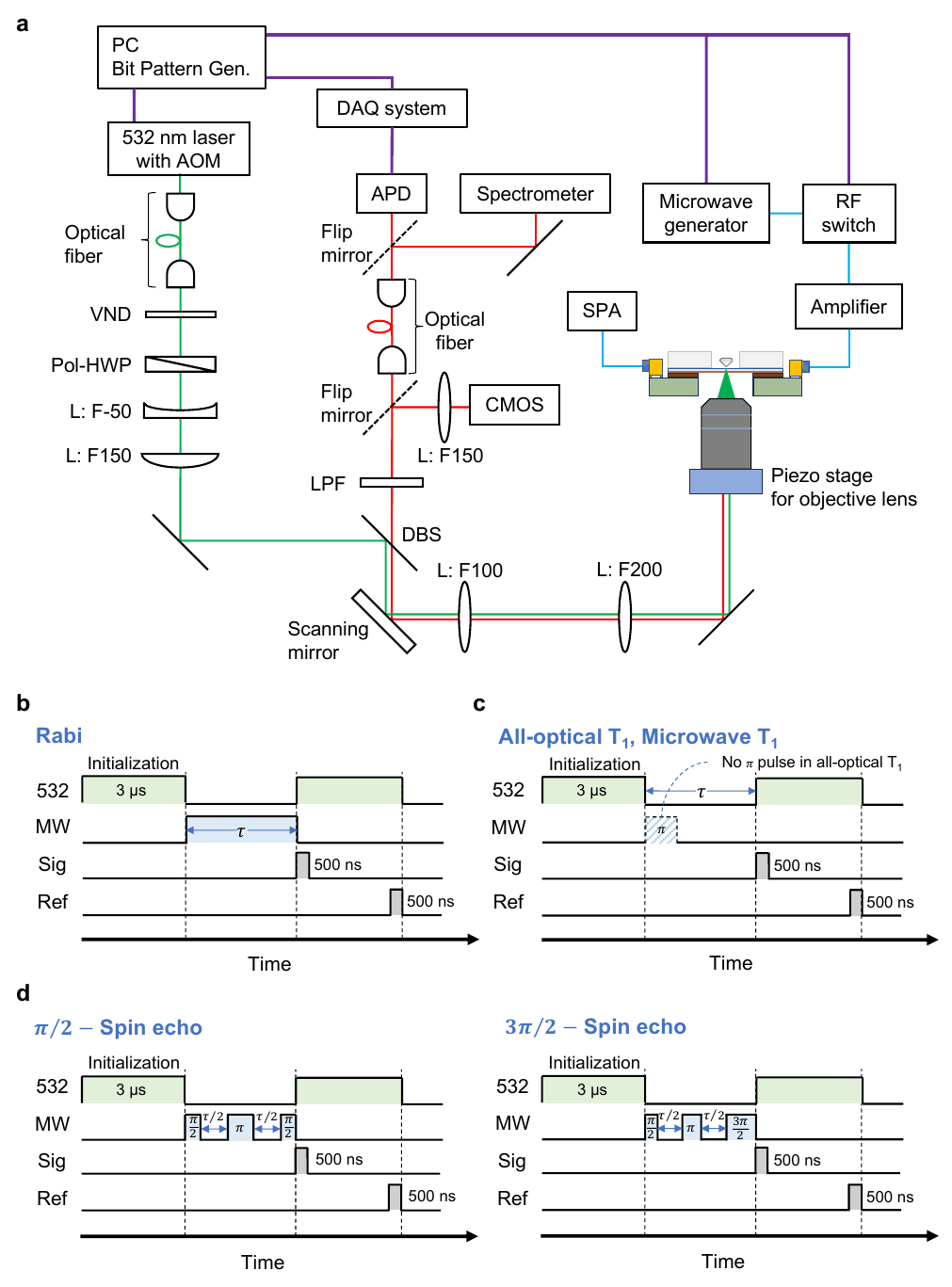}
    \caption{(a) Schematic of the ODMR setup comprising confocal fluorescence microscope and microwave excitation system. 
    AOM: acousto-optic modulator. VND: variable neutral density filter. Pol-HWP: polarizer and half-wave plate. L: lens. DBS: dichroic beam splitter. LPF: long pass filter. CMOS: complementary metal oxide semiconductor. APD: avalanche photodiode. SPA: spectrum analyzer. RF: radio frequency. 
    Schematic pulse sequences for (b) Rabi, (c) all-optical $T_1$, microwave $T_1$, and (d) $\pi/2$, $3\pi/2$-spin echo measurements.}
    \label{Optical setup and sequence}
\end{figure*}

\clearpage
\section{Loss/gain analysis of microwave excitation system including antenna characterization}

\begin{figure*}[h!]
    \centering
    \includegraphics[width=14cm]{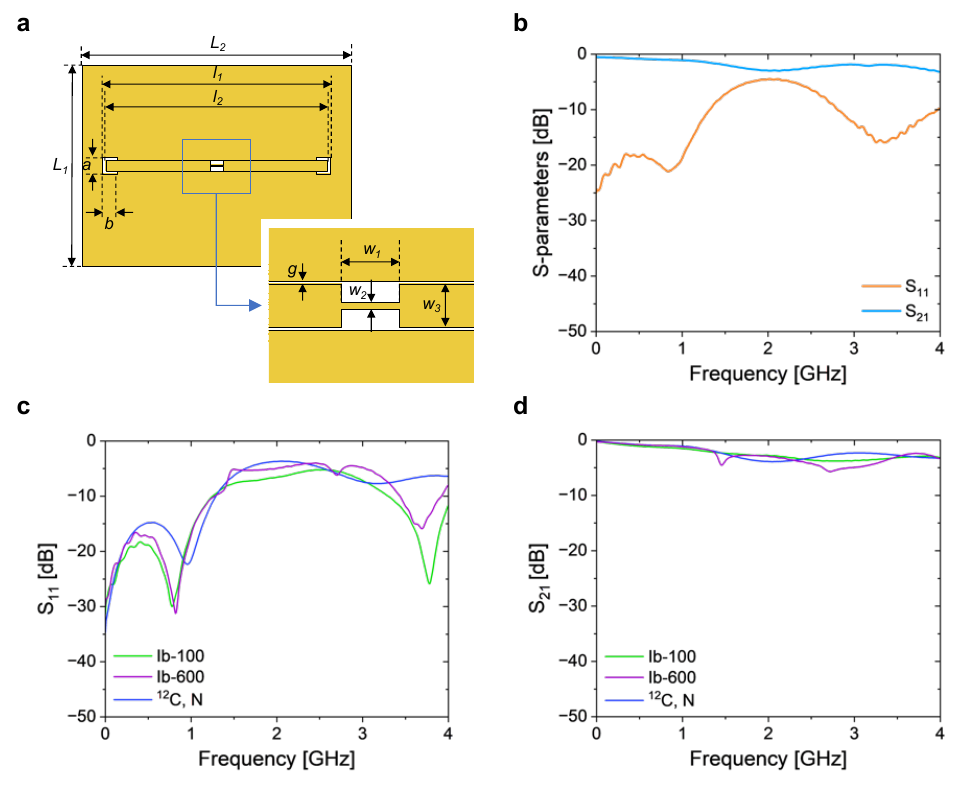}
    \caption{(a) Geometrical structure of the notch-shaped antenna patterned on the coverslip with dimensions as follows: $L_1 = 30 \ \si{\mm}$, $L_2 = 40 \ \si{\mm}$, $l_1 = 33.94 \ \si{\mm}$, $l_2 = 33 \ \si{\mm}$, $a = 2.44 \ \si{\mm}$, $b = 2.17 \ \si{\mm}$, $g = 0.10 \ \si{\mm}$,$w_1 = 2.0 \ \si{\mm}$, $w_2 = 250 \ \si{\um}$, and $w_3 = 1.5 \ \si{\mm}$. 
    (b) Experimentally measured $S_{11}$ and $S_{21}$ spectrum in the experiment of Fig.~\ref{fig2}\textbf{c}
    (c) Experimentally measured $S_{11}$ and (d) $S_{21}$ spectrum of the antenna used for the experiments of type-Ib NDs and \ce{^12C}, N-NDs presented in Figs.~\ref{fig2}\textbf{d}--\textbf{f}.    
    }
    \label{Antenna structure and S parameters}
\end{figure*}

Figures~\ref{Antenna structure and S parameters}\textbf{a, b} show the geometrical structure of a notch-shaped antenna on a coverslip and the experimentally measured S-parameters, respectively.
Table~\ref{tabel1} summarizes the losses and gains of the individual components of the microwave system in the ODMR setup. 
The microwave losses in SMA cables and RF switches were obtained from their specification sheets, and we assumed a minimum gain of $+40$ dB for the amplifier, considering the high-impedance matching between the cables and the antenna.
The loss of the antenna on the PCB at the input ($L_{R}$) and output ports ($L_{T}$) was calculated from the observed $S_{11}$ and $S_{21}$ values using the following equation ~\cite{oshimi2022glass}:
\begin{equation}
    L_{R} = 10\log_{10}{(1-10^{\frac{S_{11}}{10}})}, 
\end{equation}
\begin{equation}
     L_{T} = \frac{S_{21}}{2}.
\end{equation}
The applied microwave power was calculated from the corresponding S-parameter values (Fig.~\ref{Antenna structure and S parameters}\textbf{b}--\textbf{d}), and they were converted from dBm to mW.

\begin{table}[h!]
  \caption{List for microwave input power, loss and gain of the microwave components}
  \label{tbls1}
  \centering
  \begin{tabular}{wc{40mm}wc{30mm}wc{30mm}wc{50mm}}
	Components  & Loss or Gain [dB]  & Power [dBm]  & Notes \\ 
    \hline \hline 
    Microwave generator     &      &   $-16.0 $ & The experiment of the Fig.~\ref{fig2}\textbf{c} (bottom) \\
    RF switches 			&	$-3.5$     &   $-19.5$  & Total insertion loss	\\
    Microwave amplifier		&   $+40.0$	   &  20.5   & Amplifier gain  \\
    Microwave cables    	&	$-5.0$    	 &  15.5  & Total insertion loss \\
    Antenna reflection loss &	$-0.386$    & 15.1   &  $S_{11} = -11.0$\\
    Antenna insertion loss  &   $-0.955$    & 14.2 &  $S_{21} = -6.32$\\
     \hline 
    Total in the detection area                  &       &   $14.2$ [dBm]   & 
  \end{tabular}
  \label{tabel1}
\end{table}

\newpage
The magnetic field of the microwaves ($|\textit{\textbf{B}}|$) on the antenna in Fig.~\ref{fig2}\textbf{b} was simulated using the finite-element method (COMSOL).
Table~\ref{tabel2} summarizes the structures and corresponding materials with the relative permittivity $\epsilon$, relative permeability $\mu$, and electrical conductivity $\sigma$ used for the calculations.
The thin gold-patterned layer ($\sim 300$ nm) was regarded as an infinitely thin perfect electric conductor (PEC) in the simulations. 
A 70-mm-radius sphere surrounding the coverslip was assumed to be a perfectly matched layer with an absorption constant of $10^{-6}$ at the sphere boundary. 

\begin{table}[h!]
  \caption{List of simulation conditions}
  \label{tbls2}
  \centering
  \begin{tabular}{wc{70mm}wc{70mm}}
	Structures  & Materials  \\ 
    \hline \hline 
    Coverslip  (thickness = 0.17 mm)  & Borosilicate glass ($\epsilon = 4.6$, $\mu = 1.0$, $\sigma = 0.0~\si{S/m}$)    \\
    Thin gold layer (antenna) &	Perfect electric conductor (PEC)    \\
    70-mm-radius sphere &   Air ($\epsilon = 1.0, \mu = 1.0, \sigma = 0.0~\si{S/m}$)	  \\
     \hline 
  \end{tabular}
  \label{tabel2}
\end{table}

\clearpage
\section{ODMR spectral analysis using a double-Lorentzian function}
\begin{figure*}[h!]
    \centering
    \includegraphics[width=15cm]{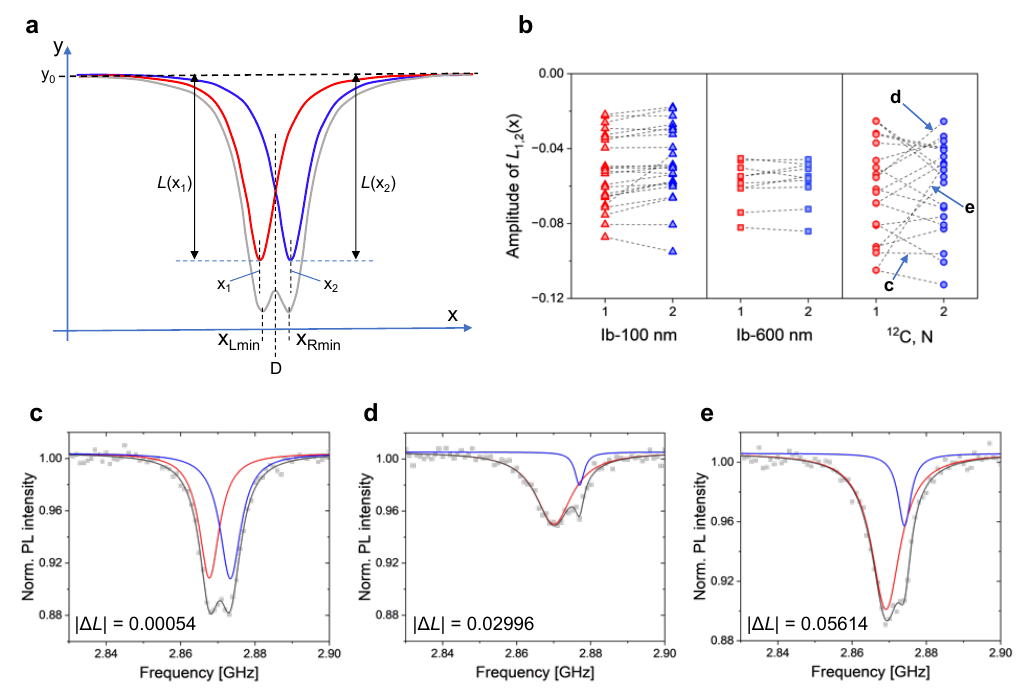}
    \caption{(a) Schematic of the double-Lorentzian function fitted to the ODMR spectra. 
    (b) Amplitudes of the fitted Lorentzian function ($L_{1} (x_{1})$, $L_{2} (x_{2})$) for the Ib-100, Ib-600, and \ce{^12C}, N-NDs. 
    \textbf{c, d, e} indicate cases for Figs.~\ref{Fitting procedure}\textbf{c, d, e}, respectively.
    (c--e) ODMR spectra with the fitting for the cases specified in Fig.~\ref{Fitting procedure}\textbf{b}.
    }
    \label{Fitting procedure}
\end{figure*}

The ODMR spectra were fitted to a double Lorentzian function comprising two Lorentzian functions (red and blue curves in Fig.~\ref{Fitting procedure}\textbf{a}): 
\begin{equation}
    \begin{aligned}
    y (x) &= y_{0} +L_1 (x) +L_2 (x), \\
    L_{1,2} (x) &= \frac{2A_{1,2}}{\pi}\frac{w_{1,2}}{4(x-x_{1,2})^2+w_{1,2}^2},
    \end{aligned}
\end{equation}
where $y_0$ is a common offset, and $A_{1,2}$, $w_{1,2}$, and $x_{1,2}$ are the peak area, line width, and peak position frequency for the left (index: 1) and right peaks (index: 2), respectively.
The ODMR depth in Fig.~\ref{fig2}\textbf{e} was obtained using $[y(x_1) + y(x_2)]/2$, and the depth in Fig.~\ref{fig2}\textbf{e} was depicted as the ``Norm.~PL~intensity''.
\textit{E} and \textit{D} shown in Figs.~\ref{fig2}\textbf{e, f}, were calculated using $E = (x_{2} - x_{1})$ and $D = (x_{1} + x_{2})/2$, respectively.
Note that the effect of the geomagnetic field ($\sim 0.5$ G)~\cite{kuwahata2020magnetometer} on \textit{E} was insignificant because its potential variation of \textit{E} is about 1.4 MHz which is comparable to the variations derived from the random NV quantization axes on the coverslip.

Figure~\ref{Fitting procedure}\textbf{b} shows the amplitudes of $L_{1} (x_{1})$ and $L_{2} (x_{2})$ in the double Lorentzian fitting for all of the measured type-Ib and \ce{^12C}, N-NDs. 
The double Lorentzian model gives successful fitting results with comparable peak amplitudes for $L_{1}$ and $L_{2}$ under the applied boundary conditions (Fig.~\ref{Fitting procedure}{c}). 
However, compared with those of type-Ib, \ce{^12C}, N-NDs sometimes exhibit asymmetric results with large differences between $L_{1} (x_{1})$ and $L_{2} (x_{2})$ ($\Delta L = |L_{1} (x_{1}) - L_{2} (x_{2})|$), which might affect the results of $E$ and $D$ in Figs.~\ref{fig2}\textbf{f,g}. 
The magnitude of $\Delta L$ does not affect the ODMR depth, as shown in Fig.~\ref{fig2}\textbf{e}, because the contrast was obtained using $ (y(x_1) + y(x_2))/2$, which only depends on the sum of $L_{1, 2}(x)$.

To evaluate this effect, we investigated individual fitting cases between NDs with small and large $\Delta L$.
We analyzed NDs with small and large $\Delta L$ values as representatives of the symmetric and asymmetric cases, respectively (Figs.~\ref{Fitting procedure}\textbf{c--e}).
To estimate the deviation of resulting from fitting destabilization (asymmetry), we compared $E$ ($D$) values determined using the above defined fitting-based calculations ($E_{\rm fit}$ or $D_{\rm fit}$) with those by manually taking the frequency positions of the two peaks of the total double Lorentzian curves: $\Delta E_{\rm man} = x_{\rm Rmin} - x_{\rm Lmin}$ or $\Delta D_{\rm man} = x_{\rm Lmin} + x_{\rm Rmin}$, where $x_{\rm L(R)min}$ indicates the minimum points of overall double Lorentzian shape (Gray curve in Fig.~\ref{Fitting procedure}\textbf{a}).
Table.~\ref{tbls3} summarizes these analyses for the three cases shown in Figs.~\ref{Fitting procedure}\textbf{c--e}.
$E_{\rm fit}-E_{\rm man}$ shows primarily positive values (0.74 and 0.71 MHz), and in the case of a large $\Delta L$, it gives a more positive value (2.18 MHz), which means that the fitting destabilization tends to overestimate the $E$ values compared to the apparent peak splitting of ODMR spectra. 
This result indicates that the observed small $E$ values for the \ce{^12C}, N-NDs could not be attributed to underestimation from the fitting destabilization; rather, the $E$ values are overestimated. 
Therefore, we conclude that the present fitting destabilization does not affect the observed tendency of small $E$ values in the \ce{^12C}, N-NDs.

$D_{\rm fit}-D_{\rm man}$ can show both negative and positive values, and the magnitude can increase up to 0.00111 GHz, which is comparable to the observed difference between \ce{^12C}, N- and type-Ib NDs. 
Therefore, fitting destabilization may result in the observed tendency of relatively large $D$ values in the \ce{^12C}, N-NDs. 
However, the ODMR spectra of ensemble NV exhibit a skirt on the low-frequency side~\cite{abrahams2023handheld}.
This low-frequency skirt is more prominent in type-Ib NDs than that in \ce{^12C}, N-NDs, and it pulls the fitting estimation of $D$ toward the lower-frequency side for type-Ib NDs. 
Thus, the relatively large $D$ values in the \ce{^12C}, N-NDs could be attributed to both fitting destabilization and intrinsic asymmetry of ODMR spectral shape in the type-Ib NDs.

\begin{table}[h!]
  \caption{Fitting destabilization for $E_{\rm fit}$ and $D_{\rm fit}$}
  \label{tbls3}
  \centering
  \begin{tabular}{wc{20mm}wc{35mm}wc{35mm}wc{35mm}}
	Case & $\Delta L$ & $E_{\rm fit}-E_{\rm man}$ [MHz]  & $D_{\rm fit}-D_{\rm man}$ [GHz] \\ 
    \hline 
    Fig.~\ref{Fitting procedure}\textbf{c} & $0.00054$ & $0.74$ & $0.00010$  \\
    Fig.~\ref{Fitting procedure}\textbf{d} & $0.02996$ & $0.71$ & $-0.00049$ \\
    Fig.~\ref{Fitting procedure}\textbf{e} & $0.05614$ & $2.18$ & $0.00111$  \\
  \end{tabular}
  \label{tabel3}
\end{table}

\clearpage
\section{Fitting procedure of the relaxation curves}

\begin{figure*}[th!]
    \centering
    \includegraphics[width=15cm]{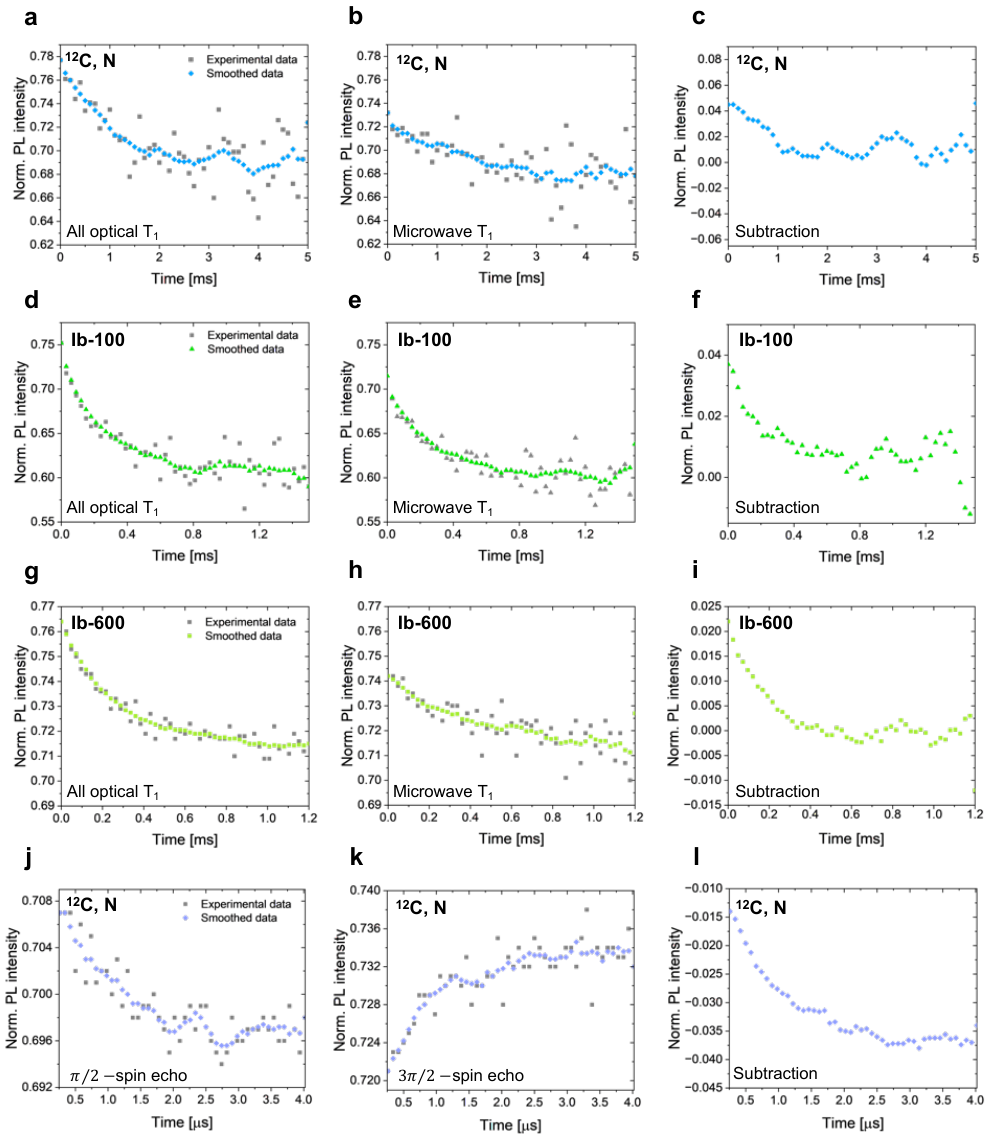}
    \caption{(a) Representative profiles of all-optical $T_1$ and (b) microwave $T_1$ relaxometry for the \ce{^12C}, N-NDs. 
    (c) The subtracted relaxation profiles.
    (Grey dots: experimental data, light blue dots: smoothed profiles by the moving average method). 
    (d)--(f) Representative profiles of all-optical $T_1$ and microwave $T_1$ relaxometry, and subtracted relaxation for the Ib-100nm NDs. (Grey dots: experimental data, green dots: smoothed profiles by the moving average method). 
    (g) Representative profiles for $\pi/2$-spin echo, and (h) $3\pi/2$-spin echo for the \ce{^12C}, N-NDs. 
    (i) The subtracted profiles. 
     (Grey dots: experimental data, lilac dot: smoothed profiles by the moving average method).
    }
    \label{12C and Ib fitting}
\end{figure*}

The raw $T_1$ relaxation profiles showed large noise, which destabilized the subsequent fitting processes.
This noise must be numerically filtered to determine $T_1$ values.
The profiles of microwave $T_1$ and all-optical $T_1$ were filtered by taking a moving average over nine data points from a total of 50 points and they were subtracted from each other, as shown in Figs.~\ref{12C and Ib fitting}\textbf{a}--\textbf{f} (subtracting the microwave $T_1$ from the all-optical $T_1$ profile). 
The subtracted profiles were fitted using bi-exponential decay ($y = y_{0} + A_{1}\exp{[-(\tau-t_{0})/t_{1}]} + A_{2}\exp{[-(\tau-t_{0})/t_{2}]}$), as defined in the Methods. 
Similarly, the two spin-echo profiles were numerically filtered to remove noise by taking a moving average over five data points with a total of 50 points. 
The filtered profiles were subtracted to obtain the final $T_2$ profiles and a stretched exponential decay function $\exp{((-\tau/T_{2})^{p})}$ was fitted to determine $T_2$. 
The value of $p$ can vary between 1.47 and 1.50 in the fitting, but we assumed $p = 1.5$ to determine $T_2$ for all $T_2$ profiles for the fitting consistency. 
Note that, in the above analysis of $T_1$ and $T_2$, the particle-based statistical error was adopted for the uncertainty of the mean relaxation times instead of the fitting error because fitting was applied to the numerically filtered data and the time step used in the measurements for the type-Ib and \ce{^12C}, N-NDs were different.

\begin{figure*}[th!]
    \centering
    \includegraphics[width=12cm]{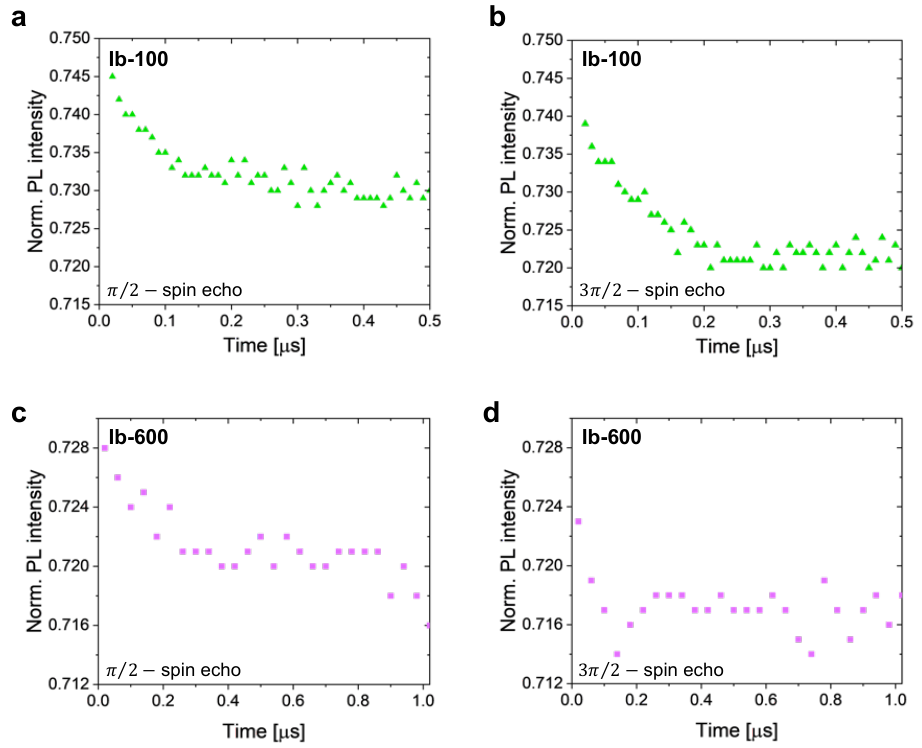}
    \caption{
    (a) Representative profiles for $\pi/2$-spin echo and (b) $3\pi/2$-spin echo for the Ib-100 (green dots: experimental data).
    (c) Representative profiles for $\pi/2$-spin echo and (d) $3\pi/2$-spin echo for the Ib-600 (purple dots: experimental data).
    }
    \label{Ib spin echo fitting}
\end{figure*}

To determine $T_2$ of the Ib-100 NDs, we fitted the stretched exponential decay function to the $\pi/2$-spin echo profiles (Fig.~\ref{Ib spin echo fitting}\textbf{a}) because the $3\pi/2$-spin echo profiles did not differ from the $\pi/2$-spin echo (Fig.~\ref{Ib spin echo fitting}\textbf{b}), presumably because of the short $T_2^*$ as compared with the microwave $\pi$ pulse (300--400 ns).
For the Ib-600 NDs, the different profiles between $\pi/2$-spin echo and $3\pi/2$-spin echo were obtained (Figs.~\ref{Ib spin echo fitting}\textbf{c, d}), but it was difficult to fit the subtracted profiles in a similar to the case of \ce{^12C}, N-NDs.
Therefore, to determine $T_2$ of the Ib-600 NDs, we fitted the stretched exponential decay function to the $\pi/2$-spin echo profiles.

\clearpage

\section{Cell culture protocol and other experimental results}

\begin{figure*}[h!]
    \centering
    \includegraphics[width=14cm]{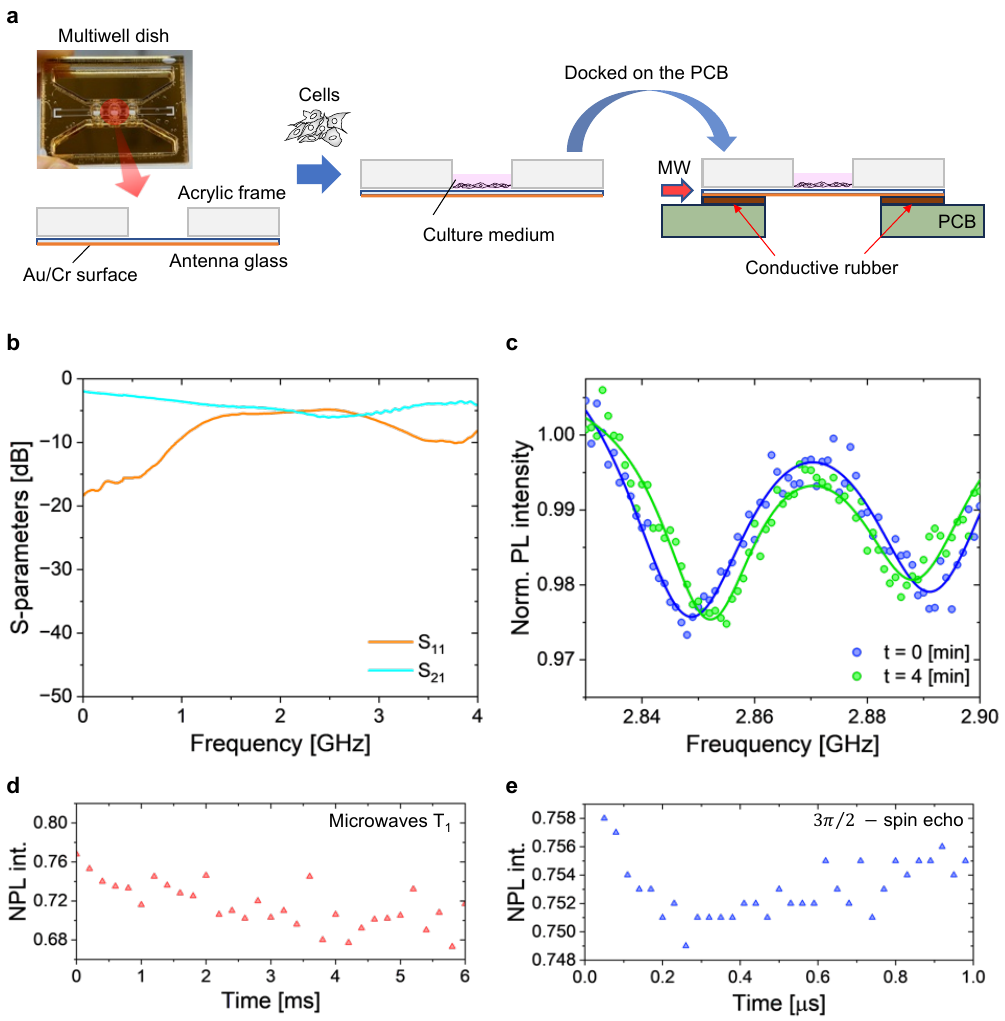}
    \caption{(a) Flow diagram for the device assembly and ODMR measurements using cells and (b) experimentally measured $S_{11}$ and $S_{21}$ spectra of the device. 
    (c) Split ODMR spectra inside cells under an external magnetic field (blue dots) and the subsequent ODMR spectra of the same NDs in 4 min (green dots).
    (d) Microwave $T_1$ and (e) $3\pi/2$-spin echo for ND2 inside the cell in Fig.~\ref{fig4}\textbf{a}.
    }
    \label{Cell culture protocol and other results}
\end{figure*}

Figure~\ref{Cell culture protocol and other results}\textbf{a} shows a flow diagram of the cell culture protocol in our home-built multiwell device with a notch-shaped microwave antenna. 
Figure~\ref{Cell culture protocol and other results}\textbf{b} shows the experimentally measured $S_{11}$ and $S_{21}$ spectra of the multiwell devices containing the cells, respectively. 
We observed that the ODMR spectral shape changed for 4 min under an external magnetic field in the CW-ODMR measurements for ND1 in the cells (Fig.~\ref{fig4}\textbf{a}), indicating the rotation of the NV orientation via rotational Brownian motion. 
This ND rotation is detrimental when measuring pulsed ODMR profiles under microwave excitation.
NDs with 100--300 nm diameters exhibit 0.1--1 kHz (1--10 ms) as the characteristic frequency (time) of the rotational Brownian motion~\cite{fujiwara2018observation}. 
For example, in the $\pi/2$-spin echo measurements, a single operation of the sequence with a single echo time finished at approximately 2.0 \si{\us} at the longest, however, this spin operation accumulated over 500 ms. 
After 500 ms of integration, the measurements proceeded to the next sequence with the next echo time. 
By sweeping the echo time from 20--1020 ns at 40 ns intervals, we obtained a single dataset for the $\pi/2$-spin echo. 
Subsequently, the dataset was collected 15 times. 
The overall measurement time was 3--5 min. 
The time-varying NV orientation lead to apparent dephasing in the final $T_2$ profiles.
Therefore, interpreting the profiles of microwave $T_1$ and two spin-echo sequences in the presence of rotational Brownian motion was difficult.

\clearpage
\section{Determination of nitrogen concentration for type-Ib NDs from FTIR spectrum }
\begin{figure*}[th!]
    \centering
    \includegraphics[scale=1]{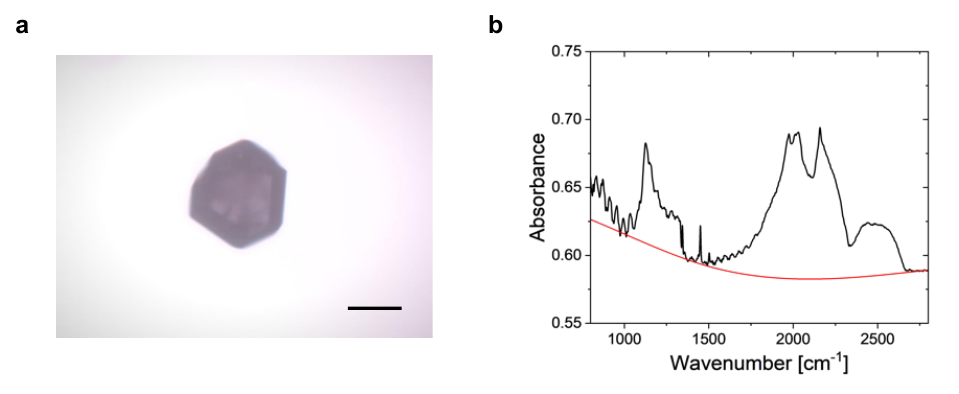}
    \caption{
    (a) Microscope image of type-Ib micro-sized diamond in the FTIR measurements.
    Scale bar: 100\,\si{\um}. 
    (b) Representative FTIR spectrum for type-Ib diamonds.
    Red curve: base line.
    }
    \label{FTIR result}
\end{figure*}

SIMS is widely used to determine the nitrogen impurity concentration in bulk diamonds~\cite{ikeda2016optical}. 
However, in our case, the crystal size of the type-Ib NDs was too small to be measured using SIMS.
EPR spectroscopy can quantify the nitrogen concentration in NDs~\cite{teraji2023nitrogen}.
However, it requires gram level amount of NDs, which is prohibitively high for the present case. 
Therefore, we employed FTIR spectroscopy to analyze the concentration of nitrogen in the conventional type-Ib NDs.
This measurement was carried out using a Jasco FTIR6200-IRT7000 micro Fourier-transform spectrometer with a KBr/Ge beam splitter and ceramic light source (see Methods). 
Figure~\ref{FTIR result}\textbf{a} shows a diamond micro-crystal taken from the same product line of the type-Ib NDs used in this study.
Figure~\ref{FTIR result}\textbf{b} shows a typical FTIR spectrum of these diamond micro-crystals. 
From this spectrum, we calculated $(\mu_{1130 \si{cm^{-1}}}/\mu_{2120 \si{cm^{-1}}}) \times 5.5 \times 25$ to determine the nitrogen concentration [N] (ppm)~\cite{liang2005synthesis}.
The [N] of these micro-crystals was estimated to 300--540 ppm, which was obtained by measuring five crystals in total.
The variation in [N] may be caused by several factors, including particulate inhomogeneity, imperfect perpendicular incidence of the NIR beams on the diamond surface because of the unavoidable angle created on the copper mesh during the FTIR measurements.

\clearpage
\section{Rabi oscillation comparison and Ramsey experiments}

\begin{figure*}[th!]
    \centering
    \includegraphics[width=13cm]{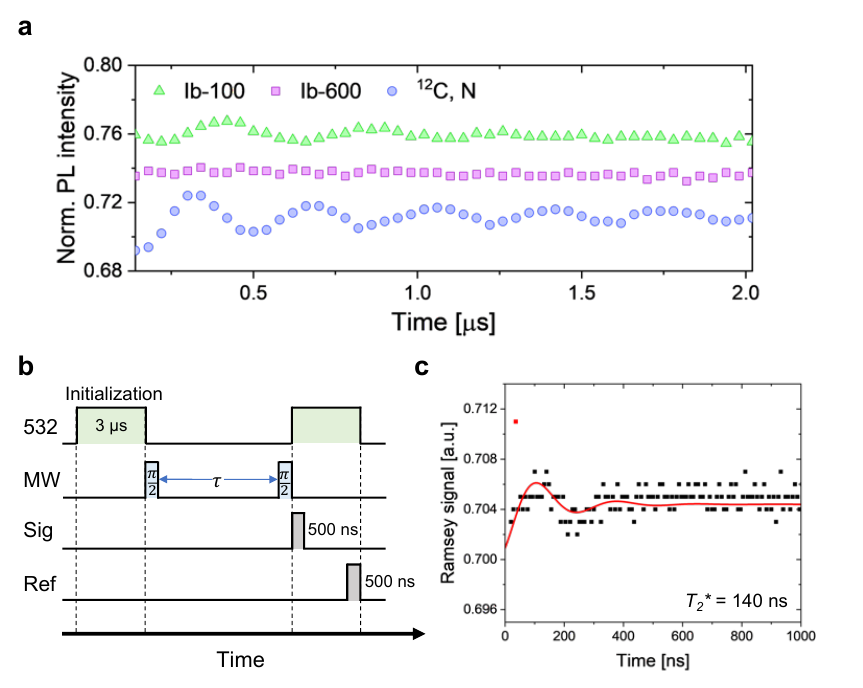}
    \caption{
    (a) Comparison of Rabi oscillations between the different types of NDs at the same microwave power.
    Green triangle: Ib-100, purple square: Ib-600, blue circle:  \ce{^12C}, N-NDs.
    In these measurements, we accumulated the data 20 times with the same measurement time step.
    (b) Schematic sequences of the Ramsey measurement. 
    (c) Ramsey oscillation of a \ce{^12C}, N-ND, where the data was accumulated 20 times. 
    The observed signal (plot data) is fitted using the  sine-damping equation (red line). A red point contained one-time noise spike and excluded from the fitting.
    }
    \label{Ramsey}
\end{figure*}

The inhomogeneity in the NV spin environment diminishes Rabi contrast (Fig.~\ref{Ramsey}\textbf{a}), which makes the subsequent pulsed ODMR measurements challenging particularly for the type-Ib NDs. 
This inhomogeneity can be evaluated via $T_{2}^*$ measured by Ramsey experiments (Fig.~\ref{Ramsey}\textbf{b}).
Ramsey oscillations for a certain \ce{^12C}, N-ND is shown in Fig.~\ref{Ramsey}\textbf{c}.

\bibliography{sumitomo}

\end{document}